\documentclass[useAMS,usenatbib]{mn2e}

\usepackage{graphicx}
\usepackage{times}
\usepackage{natbib}
\usepackage{amsmath}
\usepackage{hyperref}

\renewcommand{\d}{\mathrm{d}}
\newcommand{\gtrsim}{\ga}
\newcommand{\lesssim}{\la}

\title[Gravitational enrichment]
{The interplay between chemical and mechanical feedback from the first generation of stars}
\author[Umberto Maio et al.]{
Umberto~Maio$^{1}$\thanks{E-mail: umaio@mpe.mpg.de},
Sadegh~Khochfar$^1$,
Jarrett~L.~Johnson$^1$, and
Benedetta~Ciardi$^2$\\
${^1}$
Max-Planck-Institut f\"ur extraterrestrische Physik,
Giessenbachstra{\ss}e 1, D-85748 Garching bei M\"unchen, Germany\\
${^2}$
Max-Planck-Institut f\"ur Astrophysik Physik,
Karl-Schwarzschild-Stra{\ss}e 1, D-85748 Garching bei M\"unchen, Germany
}

\begin{document}

\date{...(draft)}
\pagerange{\pageref{firstpage}--\pageref{lastpage}}\pubyear{}
\maketitle
\label{firstpage}


\begin{abstract}
We study cosmological simulations of early structure formation, including non-equilibrium molecular chemistry, metal pollution from stellar evolution, transition from population III (popIII) to population II (popII) star formation, regulated by a given critical metallicity, and feedback effects.
We perform analyses of the properties of the gas, and use the popIII and popII populations as tracers of the metallicity.
This allows us to investigate the properties of early metal spreading from the different stellar populations and its interplay with pristine molecular gas, in terms of the initial mass function and  critical metallicity.
We find that, independently of the details about popIII modeling, after the onset of star formation, regions enriched below the critical level are mostly found in isolated environments, while popII star formation regions are much more clumped.
Typical star forming haloes, at $z\sim 15-10$, with masses between $\sim~10^7-10^8 \,\rm M_\odot$, show average SN driven outflow rates of up to $\sim~10^{-4} \,\rm M_\odot/yr$ in enriched gas, initially leaving the original star formation regions almost devoid of metals.
The polluted material, which is gravitationally incorporated in over-dense environments on timescales of $\sim 10^7\,\rm yr$, is mostly coming from external, nearby star forming sites (``gravitational enrichment'').
In parallel, the pristine-gas inflow rates are some orders of magnitudes larger, between $\sim~10^{-3}-10^{-1}\,\rm M_\odot/yr$. However, thermal feedback from SN destroys molecules within the pristine gas hindering its ability to cool and to condense  into high-density star forming regions. Only the polluted material incorporated via gravitational enrichment can continue to cool by atomic fine-structure transitions on short time scales, short enough to end the initial popIII regime within less than $10^8\,\rm yr$.
Moreover, the interplay between the pristine, cold, infalling gas and the ejected, hot, metal-rich gas leads to turbulent Reynolds numbers of the order of $\sim 10^8-10^{10}$, and contributes to the suppression of pristine inflow rates into the densest, star forming regions.
\end{abstract}




\section{Introduction}\label{sect:introduction}
One of the most outstanding problems in Astrophysics is the study of the birth and evolution of cosmic structures, from early times to present days.
Our basic understanding relies on the observations of a Universe expanding at a present-day rate ($H_0$) of $\sim 70\rm km/s/Mpc$, in which the baryonic content ($\Omega_{0b}$) is only a small fraction ($\sim 4\%$).
A form of unknown ``dark'' matter represents, instead, the bulk of the matter content ($\Omega_{0m}$), with roughly $\sim 23\%$ the total energy density of the Universe. The remaining $\sim 70\%$ is accounted for by the contribution of the so-called ``cosmological constant'', $\Lambda$, or ``dark energy'' ($\Omega_{0\Lambda}$).
Recent determinations of the cosmological parameters \cite[e.g.][]{wmap7_2010} suggest $H_0=70\rm~km/s/Mpc$ ($h=0.70$ if expressed in units of $\rm 100~km/s/Mpc$), $\Omega_{0m}=0.272$, $\Omega_{0b}=0.0456$, and $\Omega_{0\Lambda}=0.728$.
The power spectrum of the primordial perturbations follows an almost linear scaling, with an index $n=0.96$ and a normalization via mass variance within 8~Mpc/$h$ radius $\sigma_8=0.8$.
As a reference, it is common to define the standard $\Lambda$CDM model as the one with the following parameters:
$H_0=\rm 70~km/s/Mpc$,
$\Omega_{0m}=0.3$, $\Omega_{0b}=0.04$,
$\Omega_{0\Lambda}=0.7$,
$\Omega_{0tot}=1.0$,
$\sigma_8=0.9$,
$n=1$.
Stars and galaxies are thought to be born in such a cosmological setting \cite[e.g.][]{GunnGott1972,Peebles1974}.
It is believed that the first stars have formed from collapse of primordial, metal-free gas \cite[e.g.][]{GalliPalla1998,Bromm_et_al_1999,ABN2000,ABN2002,OSheaNorman2007,Yoshida2008,CenRiquelme2008} and they have subsequently polluted the surrounding medium by ejecting the metals produced in their cores.
This process is responsible for the cosmic metal pollution and the appearance of the first metals in the Universe.
In addition, the modalities of the following star formation are altered by the enhanced cooling capabilities of metal-enriched gas.
Indeed, in a primordial environment the only relevant coolants are H, He and their derived molecules  \cite[][]{SaslawZipoy1967}, like H$_2$, HD or HeH.
They are able to cool the medium only down to temperatures that result in Jeans-masses  $\sim 10-100\,\rm M_\odot$, and therefore present-day, common $\rm\sim M_\odot$ stars cannot be born.
Metal or dust cooling \cite[e.g.][]{Omukai2005,Dwek2007,Schneider2010,Grassi_et_al_2010arXiv,DwekCherchneff2011} and turbulence effects at the moment when the baryons become self-gravitating \cite[e.g.][]{Clark_et_al_2010} are thought to be able to lower these limits down to solar scales, but within a large mass range.
\\
Thus, the very first (population III, hereafter popIII) stars are supposed to be \cite[e.g.][]{WW1995,Schaerer2002} massive, and short-lived (up to $\sim 10^6\,\rm yr$), while the following (population II, hereafter popII) ones have low mass (solar-like) and can live much longer (up to $10^9-10^{10}\,\rm yr$).
Cosmological simulations addressing the role of metal enrichment in the history of the Universe have been performed, both by adopting standard recipes for star formation\footnote{
Several studies use a different way to deal with star formation, i.e. via ``sink'' particles, but we warn the reader that very recent calculations \cite[][]{Bate_et_al_2010} have shown two main limitations of this approach: overprediction of the accretion rates, and mispredictions of the particle orbits.
}
\cite[e.g.][]{KatzGunn1991,Cen1992,CenOstriker1992,Katz1992,Katz1996,SpringelHernquist2003} and by following the detailed numerical network to account for molecular evolution and early, catastrophic run-away collapse \cite[e.g.][]{Gnedin1998,Yoshida_et_al_2003,Bromm_et_al_2003,RicottiOstriker2004,Wise2008,Greif_et_al_2010,Maio2006,Maio2007,Maio2010}.
Recent numerical simulations including both popIII and popII yields \cite[][]{Tornatore2007} and, additionally, full molecular evolution \cite[][]{Maio2010} seem to suggest that the {\it transition} between the two regimes is expected to take place after the very first metal-pollution events (dominated mainly by O and C).
Due to the high metal yields of primordial SN explosions, average metallicities reach $\sim 10^{-3}-10^{-2}Z_\odot$ on very short time scales \cite[][]{Maio2010}, and the subsequent phases of the cosmic star formation history are dominated by the population II evolution with just minor, residual contributions from the population III regime \cite[e.g.][]{Tornatore2007,Maio2010}.\\
{
Also the impacts of primordial supersonic gas bulk flows \cite[][]{TseliakhovichHirata2010,Tseliakov_et_al_2010arXiv} with respect to the underlying dark-matter distributions were recently studied in numerical simulations for the first time by \cite{Maio2010b}, followed by \cite{Stacy_et_al_2010arXiv}, and the conclusion was that early structure formation is suppressed, with a cascading effect inducing a time-delay in gas collapse of some $10^7\,\rm yr$.
}\\
The detectability of the first stars has been widely discussed in the literature \cite[for a complete review, see][]{CiardiFerrara2005}, and it is generally believed that next-generation telescopes might be able to observe the first galaxies at high redshift \cite[e.g.][]{WeinmannLilly2005,Crowther_et_al_2010,Johnson_et_al_2010,DijkstraWyithe2010}.
An important point to stress is our ignorance about the detailed metal-pollution history at early times, when star forming haloes are ``small'' (i.e. with maximum masses of the order of $10^5-10^6\,\rm M_\odot$ at redshift $z\sim 20$, and $\sim 10^8- 10^9\,\rm M_\odot$ at $z\sim 10$).
It is reasonable that such small haloes are not able to retain all their gas, given the impact from supernova feedback \cite[e.g.][]{Aguirre_et_al_2001,SpringelHernquist2003,Whalen_et_al_2008}.
Another issue is the behavior of pristine gas in the presence of metal-rich gas.
Detailed studies of phase mixing are still needed to address this question rigorously \cite[see e.g.][]{Spitzer1962,CowieMcKee1977,Brookshaw1985,Sarazin1988,Monaghan1992,ClearyMonaghan1999,AvillezMacLow2002,KlessenLin2003,Jubelgas_et_al_2004,Monaghan_et_al_2005,Wadsley_et_al_2008,Greif2009,Shen_et_al_2010}.
\\
In this paper, we will outline the main features of the collapse of H$_{\rm 2}$-cooled pristine gas and the subsequent pollution from early structure formation; in particular, we will address the impact of metal-rich gas on pristine, molecular gas and the interplay between the popIII and the popII regime.
The problems we want to address are:
is early metal pollution efficient in enriching the surrounding gas?
How much metal mass is expelled from the star formation sites?
How much metal mass is trapped in dense, star forming environments?
On which timescales?
Are metals coming from nearby star forming regions or are they re-incorporated by the same halo where they were produced?
After the first star formation episodes, can the pristine, molecular gas inflows restore the popIII regime?
Which are the preferred popIII and popII star formation sites?
How do different popIII IMF's influence the picture?
\\
This work is divided as follows:
in Sect. \ref{sect:methods}, we describe the simulations used and outline the method adopted, in Sect. \ref{sect:results}, we present our results, and in Sect. \ref{sect:disc}, we discuss and conclude.


\section{Methods}\label{sect:methods}

We use the N-body/SPH numerical simulations presented in \cite{Maio2010}, and performed by using the entropy-conserving code Gadget \cite[][]{Springel2005}.
The code contains a complete set of the most relevant, non-equilibrium, chemical reactions (involving H, H$^+$, H$^-$, He, He$^+$, He$^{++}$, D, D$^+$, HD, H$_2$, H$_2^+$, HeH$^+$, e$^-$), in order to properly follow the early stages of gas cooling \cite[e.g.][and references therein]{Yoshida_et_al_2003,Maio2006,Maio2007,Maio2009,Maio2010}.
We also include several metal species produced from stellar evolution (C, O, Si, Fe, Mg, S, Ca), according to two different populations \cite[][]{Tornatore2007}: the popIII stars, described by a top-heavy IMF\footnote{
The exact slope for the popIII IMF has little effect on the  metallicity evolution. The main contribution to the metallicity is via pair-instability supernovae (PISN) whose fraction only show slight changes between $\sim 0.35-0.40$, for variations of the slope within the limits of $\sim $[-3.0, -0.5], given an IMF range  of $\sim$ [100~M$_\odot$, 500~M$_\odot$] \cite[][]{Maio2010}.
},
and the popII stars, described by a common Salpeter IMF.
The transition between the two regimes is determined by the critical metallicity, $Z_{crit}$ \cite[see also][]{Schneider2003,BrommLoeb2003}.
We consider several scenarios for the possible popIII IMF, by adopting different IMF mass ranges, slopes, yields, and critical metallicities. In the cases where a top-heavy IMF was assumed \cite[see also e.g.][]{RicottiOstriker2004,Karlsson2008,Maio2010}, the pollution was faster and stronger, reaching $\sim 10^{-3}-10^{-2}\,Z_\odot$ immediatelly after the onset of star formation.
The cases of moderate- or low-mass IMF show a similar, but more gradual, trend.
Since the goal of this paper is an investigation of the dynamical properties of the pollution process, without loss of generality, we will focus mainly on the top-heavy IMF runs, and we will only refer to the other runs as comparison.
We consider different simulations in the framework of the cosmological standard model, assuming $Z_{crit}=10^{-3}, 10^{-4}, 10^{-5}, 10^{-6} Z_\odot$, using  high-resolution boxes of $1\,\rm Mpc$ a side, and $2\times 320^3$ particles for gas and dark-matter.\\
The main processes which cause pristine gas to cool are particle collisions and molecule creation, with typical molecule fractions becoming larger than $\sim 10^{-2}$ and completely dominating early gas cooling.
At the bottom of the cooling branch, we assume that star formation and feedback processes occur \cite[e.g.][]{SpringelHernquist2003, Maio2009}, and metals are produced from stellar evolution.
The feedback processes we deal with are thermal feedback, mechanical feedback (both shocks and winds) and the chemical feedback: thermal feedback from supernov{\ae} injects entropy into the medium heating up closeby particles and destroying molecules;
kinetic wind feedback is stochastically taken into account, according to \cite{SpringelHernquist2003}, and it is responsible for removing away from the collapsing sites and spreading around gas and metals at velocities of the order of some $\sim 10^2\,\rm km/s$;
chemical feedback rules the actual stellar IMF and leads the transition from the primordial popIII regime to the following popII one.
In the popIII star formation regime, the only stars contributing to metal pollution are the ones in the mass range [140~M$_\odot$, 260~M$_\odot$], that explode as PISN after a life-time of at most $\sim 10^6 \,\rm yr$.
Their contribution to the total SFR rapidly drops down $\sim 10^{-3}-10^{-4}$  \cite[see][]{Maio2010}, and this seems consistent with what can be inferred from more recent observational analyses \cite[e.g.][ and private communication]{Papovich_et_al_2010}.
Given the many uncertainties on the primordial IMF \cite[e.g. see][]{Yoshida2006,Yoshida_et_al_2007,CampbellLattanzio2008,SudaFujimoto2010}, we also consider the other limiting case of a Salpeter-like popIII IMF, with mass range similar to popII, but with metal yields consistent with those from metal-free stars \cite[see Sect. 4.1. in][for further details]{Maio2010}.
\\
In the case of popII, the surrounding environment is polluted via SNII and SNIa, which occur $\sim 10^{7}-10^{8}\,\rm yr$, and $\sim 10^9-10^{10}\,\rm yr$ following the formation of their progenitor stars, respectively.
Moreover, stars can loose a significant fraction of mass during their AGB phase.
Metals are injected into the ISM and pollute it by means of stellar winds.
First star formation events take place at $z\sim 16.3$, when the Universe is roughly $2.4\times 10^8$ years old.
\\
In the following, results from our analyses will be presented and discussed.
We will refer to enriched gas particles with metallicity, $Z$, higher than or equal to $Z_{crit}$ as popII particles, while to the ones with metallicity below $Z_{crit}$ as popIII particles.
Particles with $Z=0$ are referred to as pristine particles.


\section{Results}\label{sect:results}
In this section we present our results, paying attention first to some global properties of metal enrichment (Sect. \ref{sect:ff}), then to the statistical (Sect. \ref{sect:statistics}) and dynamical (Sect. \ref{sect:dynamics}) ones. We also discuss the features and the interplay between the two different stellar populations (popIII and popII).


\subsection{Volume filling factor}\label{sect:ff}
\begin{figure*}
\centering
\includegraphics[width=0.45\textwidth]{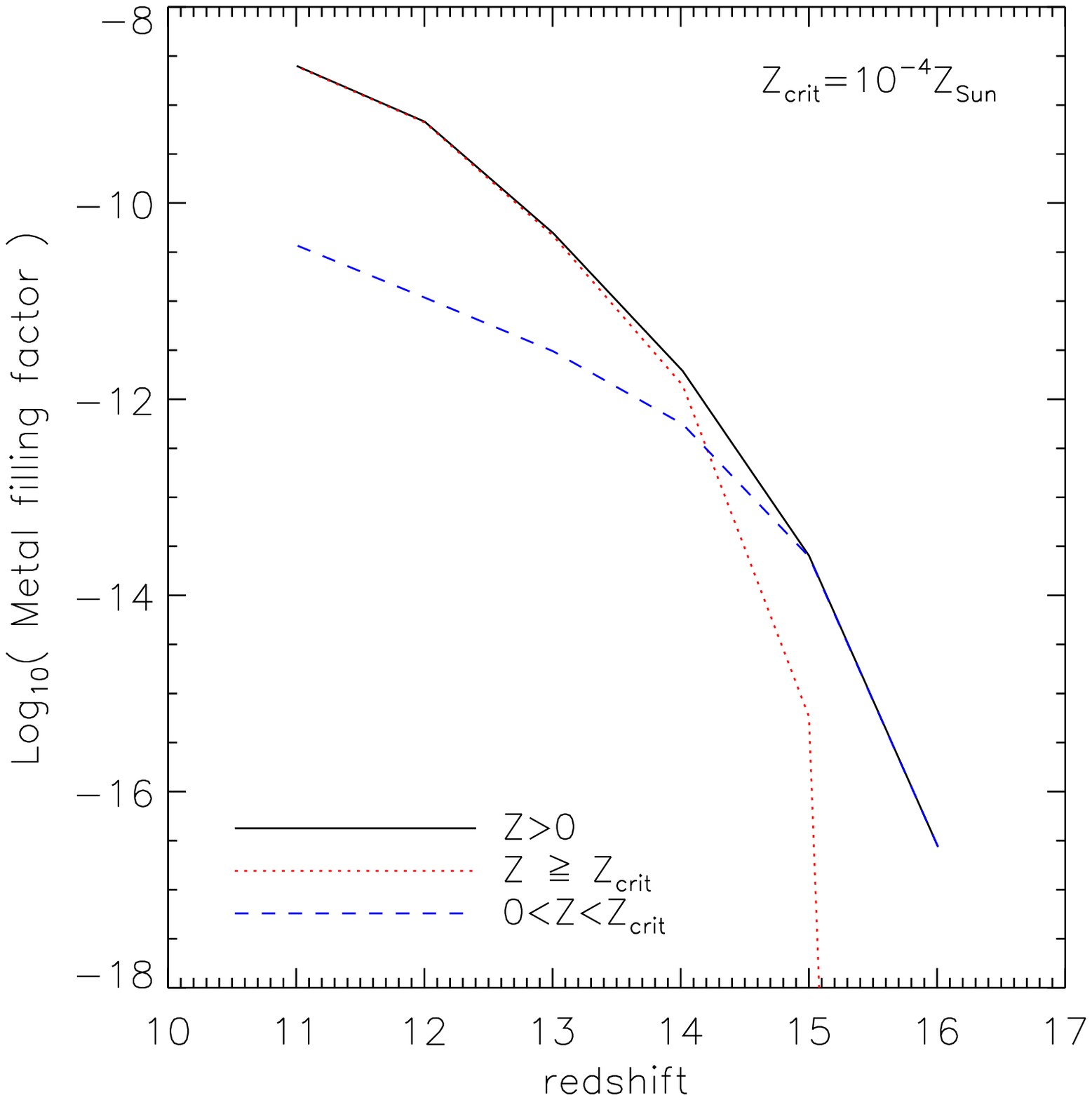}
\includegraphics[width=0.45\textwidth]{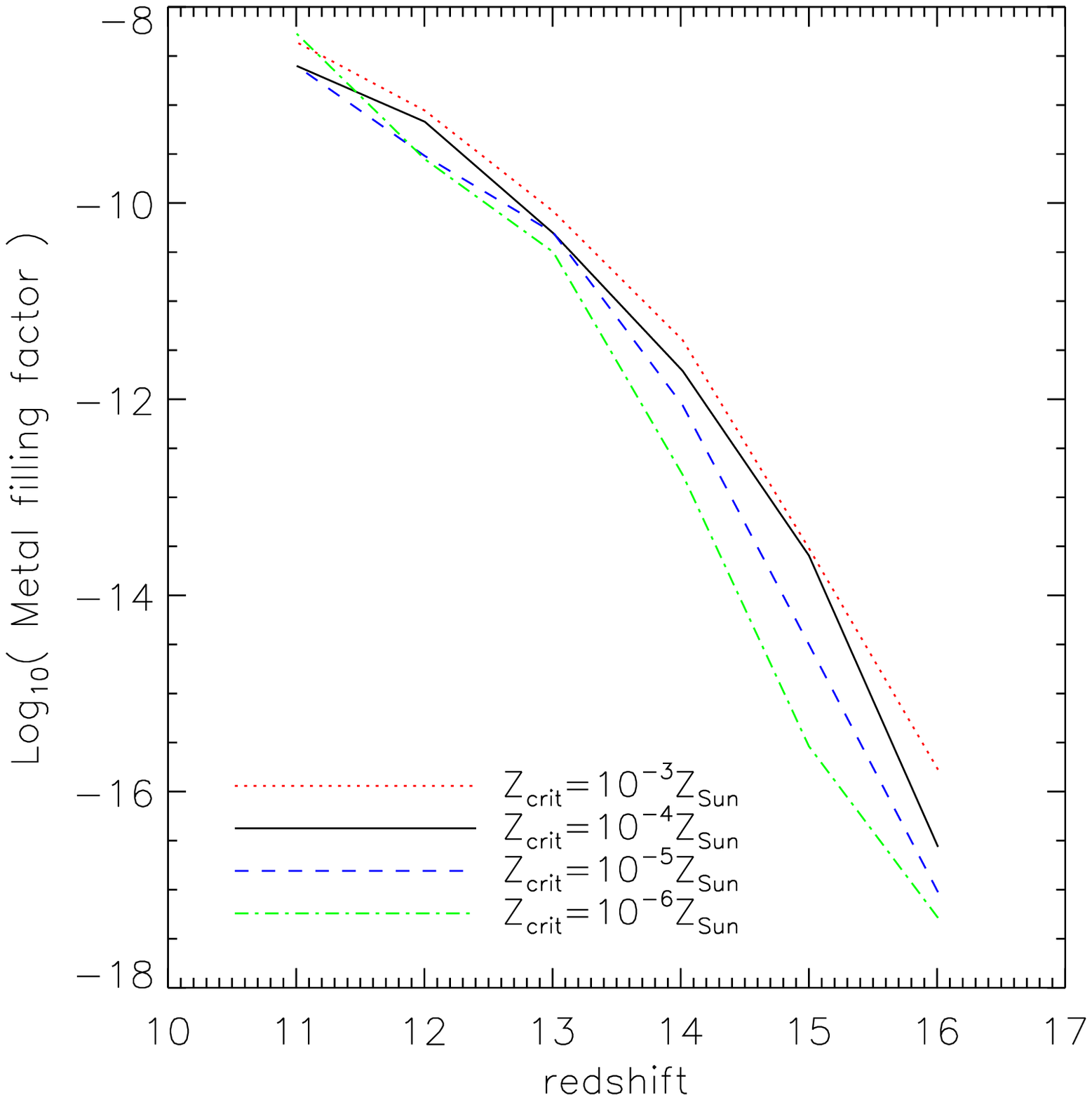}
\caption[Distribution]{\small
{\it Left panel}: metal filling factors for different regimes.
Solid, dashed, and dotted lines are the filling factors computed for particles having metallicity $Z>0$, $0<Z<Z_{crit}$, and $Z\ge Z_{crit}$ ($Z_{crit}=10^{-4}Z_\odot$), respectively.
{\it Right panel}: metal ($Z>0$) filling factors for simulations with different critical metallicity.
Solid, dotted, dashed, dot-dashed lines refer to $Z_{crit}=10^{-3}Z_\odot$, $Z_{crit}=10^{-4}Z_\odot$, $Z_{crit}=10^{-5}Z_\odot$, and $Z_{crit}=10^{-6}Z_\odot$, respectively.
}
\label{fig:ff}
\end{figure*}
To have a general idea of the metal spreading process, we start presenting the basic features of the volume filling factor.\\
In Fig. \ref{fig:ff} (left panel), the volume filling factor of metal enriched particles as a function of the redshift is shown.
The volume filling factor, $f_V$, is defined as follows\footnote{
Here, there is no difference by defining $f_V$ via physical or comoving quantities.
}:
\begin{equation}
f_V \equiv \frac{\sum_i m_{Z,i}/\rho_{Z,i}}{\sum_j m_j/ \rho_j} \sim \frac{\sum_i m_{Z,i}/\rho_{Z,i}}{V},
\end{equation}
with $i$ and $j$ integers running over the number of the enriched particles and of all the SPH particles, respectively,
$m_{Z}$ particle metal mass,
$\rho_{Z}$ metal-mass density,
$m$ particle total mass,
$\rho$ total-mass density,
and $V$ simulation box volume.
The solid line refers to the filling factor computed by considering only the polluted particles ($Z>0$), the dashed line refers to the particles with metallicity $0<Z<Z_{crit}$ (popIII regime), and the dotted line to the particles with metallicity $Z\ge Z_{crit}$ (popII regime).
In all the three cases, the general trend is similar: the filling factor increases as redshift decreases. This is simply due to ongoing metal pollution events at $z\lesssim 16$.
In more detail, at early times, some metals are produced, after the first PISN explosions, and the production sites are the very dense, clumped regions (with overdensities $\delta > 10^4$), so the resulting volume filling factor is rather small ($\sim 10^{-16}$).
At later times, particles are ejected from these dense regions and their densities rapidly decrease, which results in an increase of $f_V$.
In this regime, low-density particles, which experienced mechanical (winds) and chemical (metal pollution) feedback from stellar evolution, are dominating.
Interestingly, the discrimination by $Z_{crit}$ shows a higher filling factor for the particles with $0<Z<Z_{crit}$, at redshift $\sim 14-16$, and a behaviour dominated by the $Z\ge Z_{crit}$~--~particles, at $z\lesssim 14$.
This has to be interpreted as a consequence of the metal-enrichment process, as most of the enriched regions enter the popII regime very rapidly, while the residual popIII areas are confined to isolated regions or on the border of popII star forming sites and contribute less to $f_V$.
The exact value of the critical metallicity, $Z_{crit}$, does not alter significantly the properties of metal pollution, because of the high metal yields of early stars: larger $Z_{crit}$ slightly postpone the epoch when popII becomes dominant, while smaller $Z_{crit}$ anticipate it.
In the extreme case of $Z_{crit}=10^{-6}\,\rm Z_\odot$, then, $f_V$ is immediatelly led by the popII regime, basically since the onset of star formation.
A Salpeter-like popIII IMF will only induce a short delay in the overall behaviour, due to the longer stellar lifetimes \cite[see also][]{Maio2010}.
To illustrate the insensitivity of the volume filling factor to the adopted value of $Z_{crit}$, in the same Fig. \ref{fig:ff} (right panel) we compare the filling factors obtained for simulations having different critical metallicities: dotted, solid, dashed, dot-dashed lines refer to $Z_{crit}=10^{-3}Z_\odot$, $Z_{crit}=10^{-4}Z_\odot$, $Z_{crit}=10^{-5}Z_\odot$, and $Z_{crit}=10^{-6}Z_\odot$, respectively.
As expected, there are no strong variations among these cases, and despite some differences at the initial times, the overall trends easily converge.
The one-order of magnitude difference in the very early epoch is a consequence of the chemical feedback: higher values of $Z_{crit}$ allow the popIII regime to last longer and to pollute more with respect to the lower-$Z_{crit}$ cases.


\subsection{Statistics}\label{sect:statistics}

\begin{figure*}
\centering
\includegraphics[height=0.2\textheight, width=0.33\textwidth]{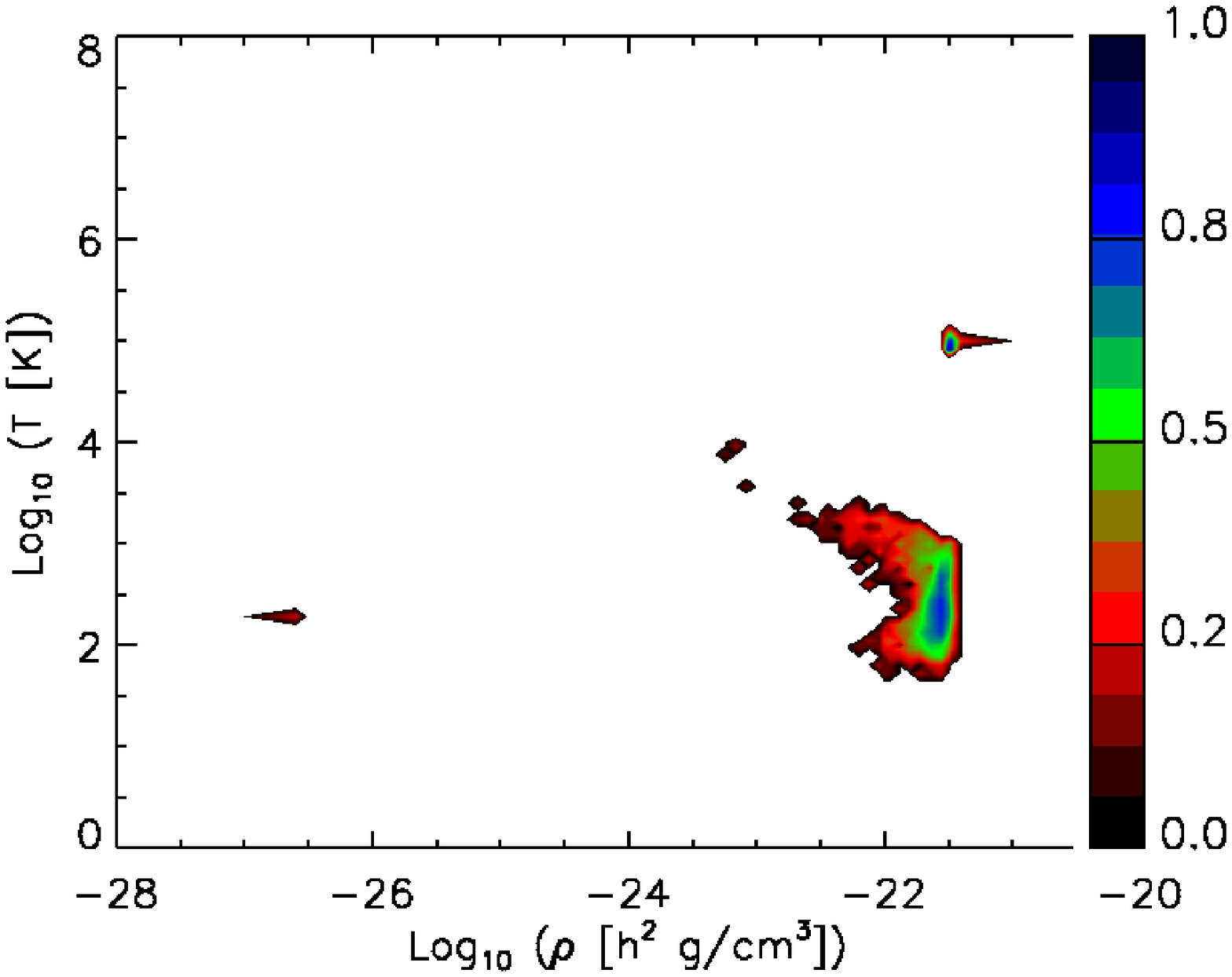}
\includegraphics[height=0.2\textheight, width=0.33\textwidth]{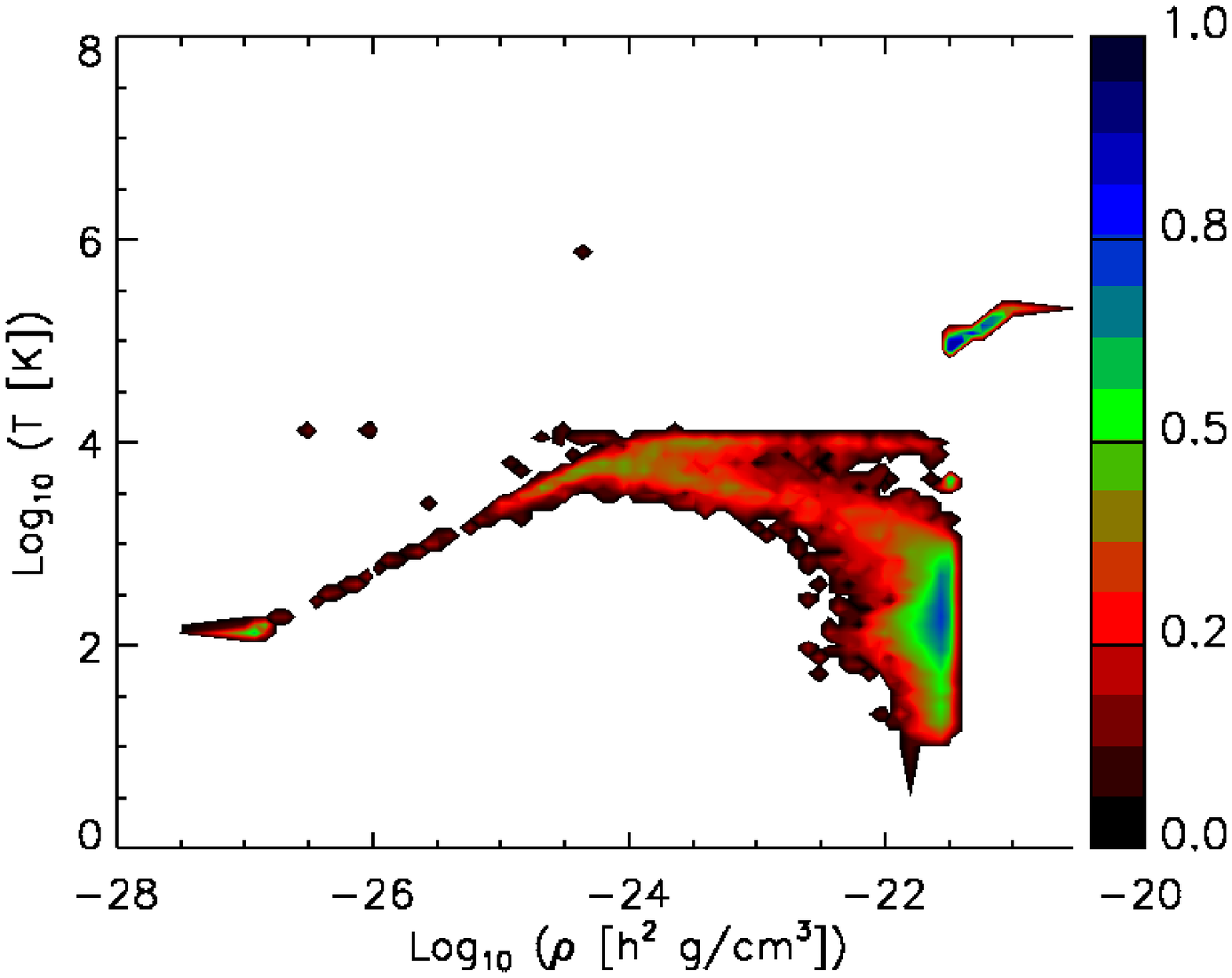}
\includegraphics[height=0.2\textheight, width=0.33\textwidth]{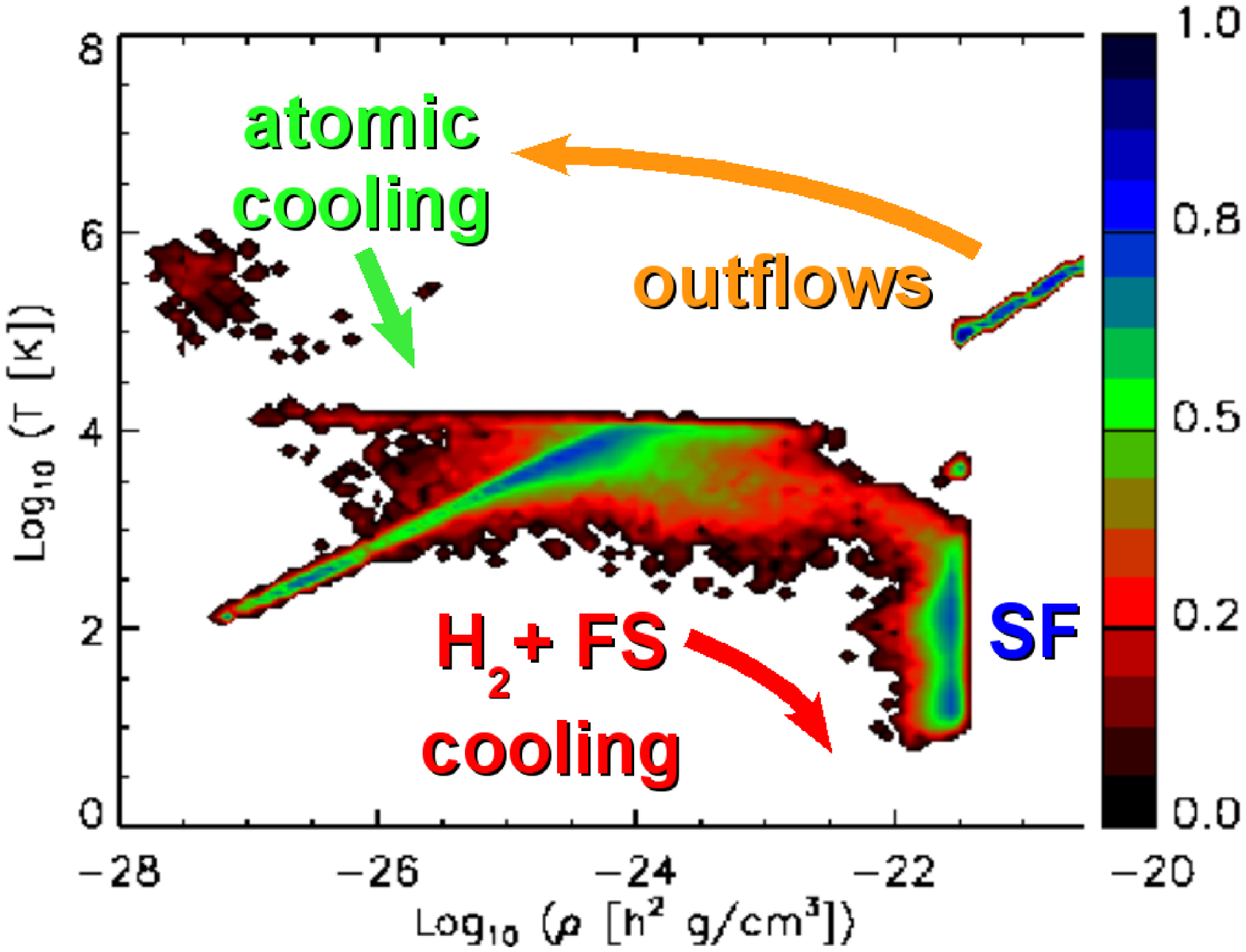}
\caption[Phase diagrams]{\small
Redshift evolution of metal-enriched particles in the $\rho-T$ diagram.
Data refer to redshift $z=15.00$ (left), $z=13.01$ (center), and $z=11.00$ (right).
The color scale corresponds to the fraction of particles in a given pixel (normalized to 1).
The critical density at $z=11$, in the standard $\Lambda$CDM model, corresponds to $\sim 10^{-26}\,h^2\rm g/cm^3$.
In the right panel, mechanical and thermodynamical evolution of the gas is shown at different stages, via different arrows (see labels): low-temperature cooling from molecules (mainly H$_2$) or fine-structure (FS) transitions leads to star formation (SF) and subsequent gas outflows into the low-density regions; there hot gas cools down to $\sim 10^4\,\rm K$ via resonant atomic transitions and below $10^4\,\rm K$ via molecular and fine-structure transitions.
}
\label{fig:PhaseEvolution}
\end{figure*}
To go deeper into the analysis, we study in detail the relevant statistical properties of the gas during the early stages of metal enrichment, and then we show the implications for the popIII and popII regimes.

\subsubsection{Global features}
In Fig. \ref{fig:PhaseEvolution}, we show the redshift evolution of metal-enriched particles in the phase diagram ($\rho-T$), at  $z=15.00$ (left), $z=13.01$ (center), and $z=11.00$ (right).
The color scale denotes the fraction of particles at any given density and temperature.
The corresponding age of the Universe is $2.66\times 10^8\rm yr$, $3.24\times 10^8\rm yr$, and $4.09\times 10^8\rm yr$.
In the plots, we immediately recognize the gas which falls in the haloes and gets shock heated (low densities and low temperatures), the isothermal phase (where the heating is couterbalanced by cooling, resulting in a roughly constant temperature of $\sim~10^4\,\rm K$), and the catastrophic cooling branch of over-dense gas (high densities and low temperatures) that enters the star forming regime as soon as it reaches a threshold of $\sim~140\,h^2\,\rm cm^{-3}$.
Above this threshold, stochastic star formation and feedback effects take place (high density and high temperatures) and lead the hot outflows from dense environments into lower-density regions (low densities and high temperatures).
What emerges from the diagrams is also the behaviour of metal enrichment after the initial pollution events.
At the beginning (left panel), only the gas at high number densities ($\sim~10^2-10^3\,h^2\rm cm^{-3}$) is affected, as this is where star formation is occurring.
As star formation and metal spreading proceed, the surrounding, lower-density regions are the next to be affected by metal pollution.
The high temperature particles ($T>10^4\,\rm K$ in the diagrams) are the ones impacted by SN feedback:
in particular, hot, dense gas has been heated, within star forming regions, by the high SN temperatures (thermal feedback), while hot, low-density gas has been ejected from star forming regions by SN winds (wind feedback).
Interestingly, the enrichment pattern of the phase diagram follows the molecular catastrophic-cooling branch, meaning that an ``inside-out'' pollution mode is taking place.
Indeed, since the critical density, in the $\Lambda$CDM model, at $z\sim 11$, is roughly $10^{-26}\,h^2\rm g/cm^3$, Fig. \ref{fig:PhaseEvolution} suggests that the very early enrichment process affects the collapsed objects\footnote{
According to the ``top-hat'' model, collapsed objects are the ones with densities of the order of $\sim 200$ times the critical one.
}
by moving from the center towards the borders and the lower-density environments.
\\
\begin{figure*}
\centering
\hspace{-0.4cm}
\includegraphics[width=0.33\textwidth]{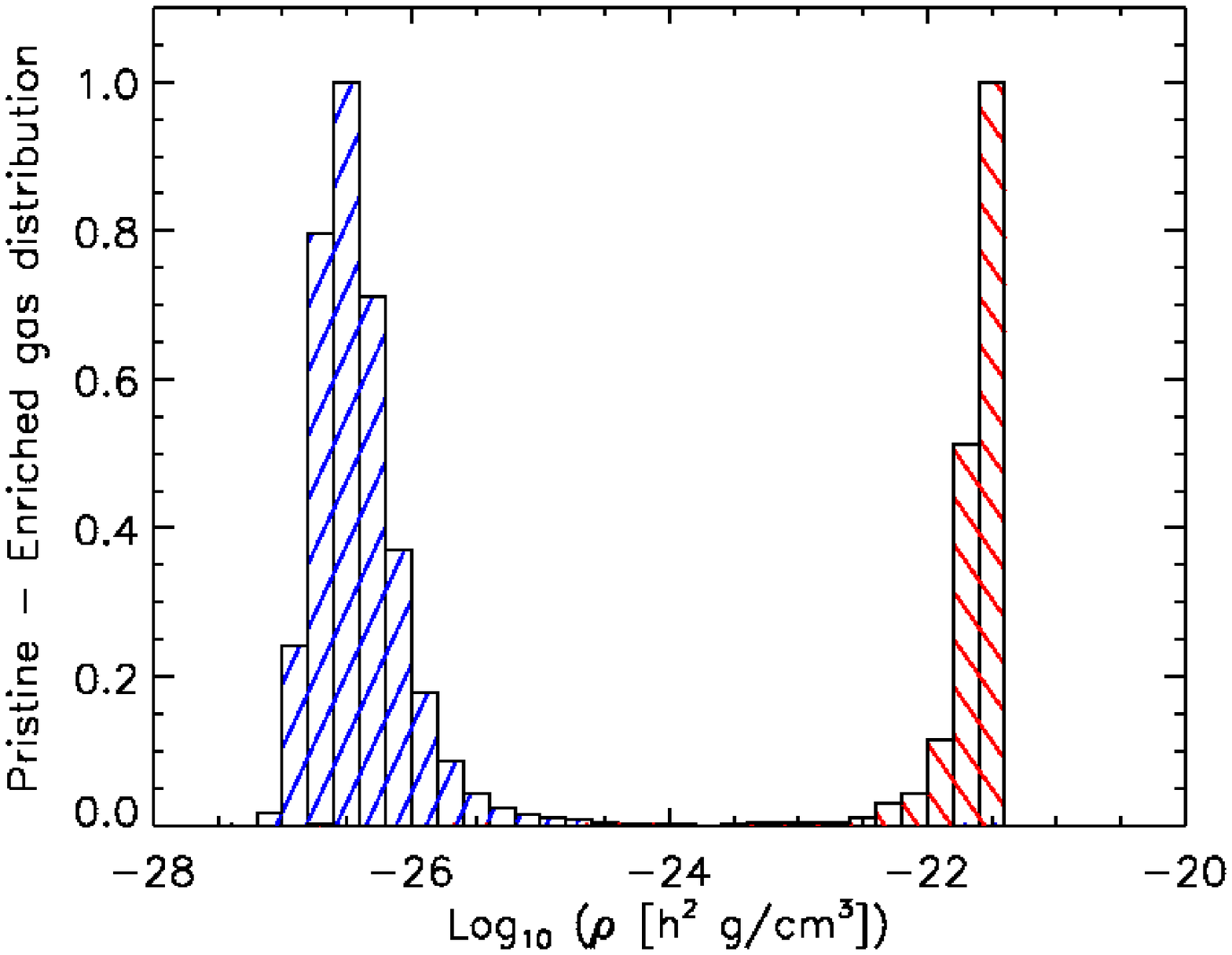}
\includegraphics[width=0.33\textwidth]{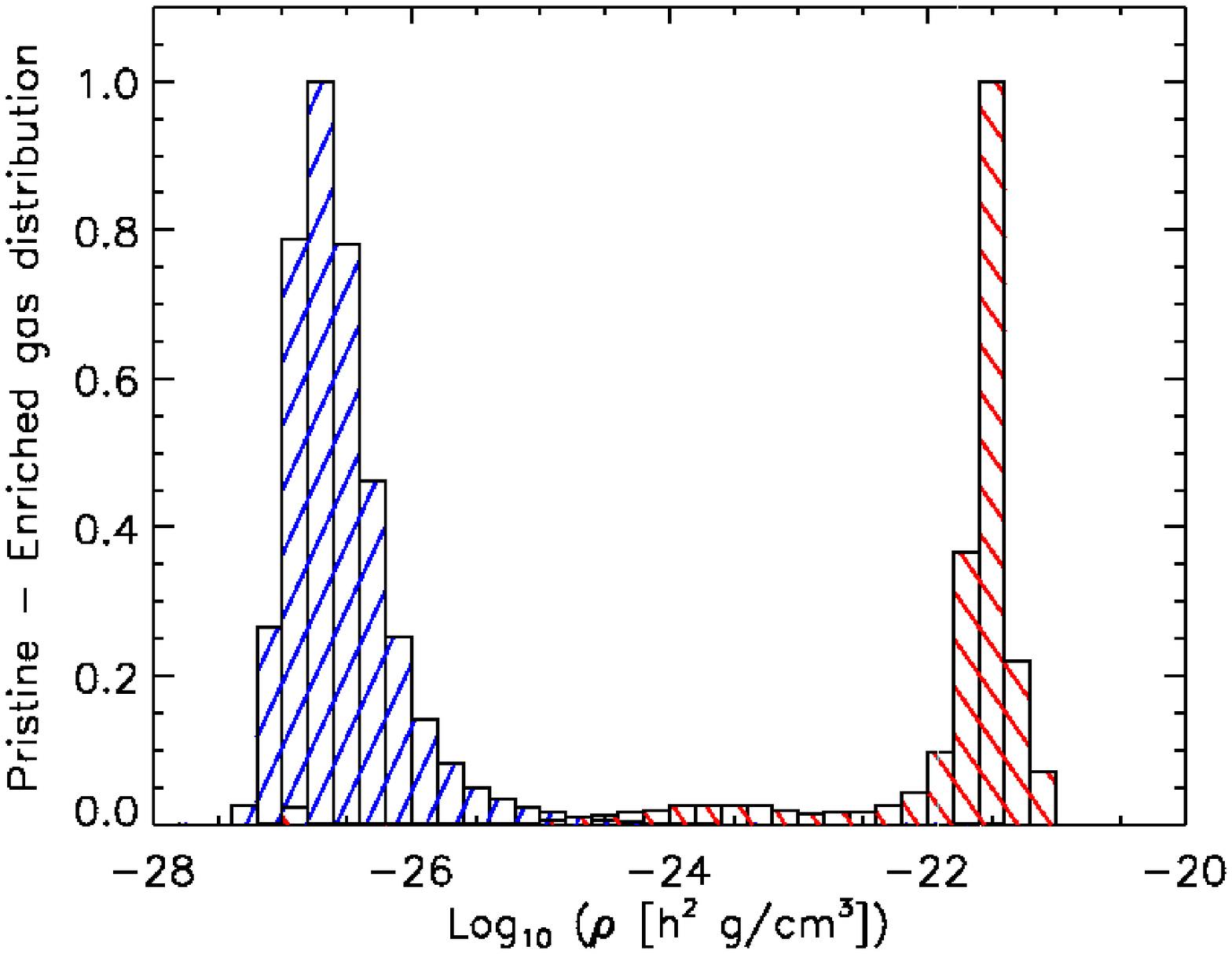}
\includegraphics[width=0.33\textwidth]{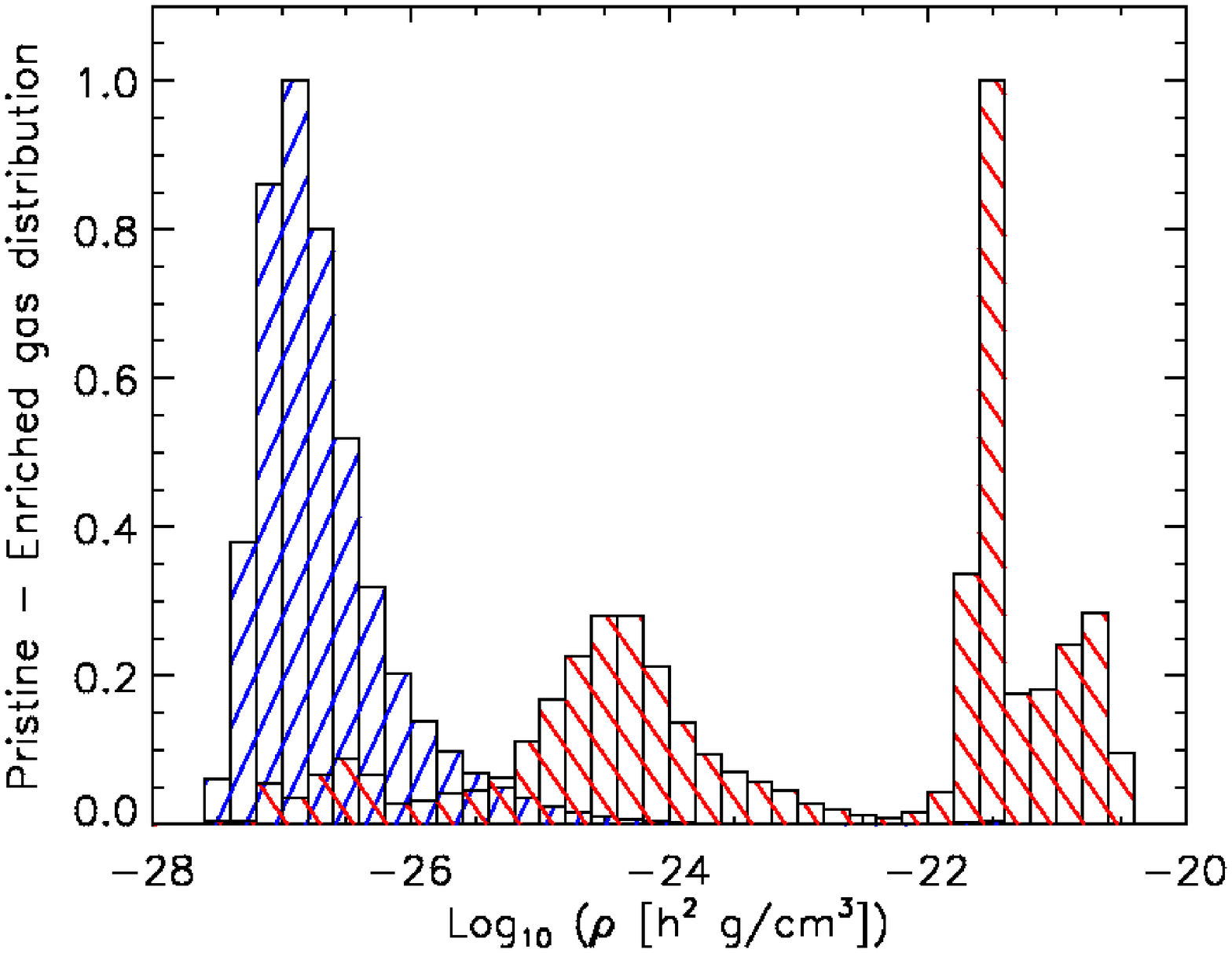}
\caption[Gas distribution]{\small
Redshift evolution of gas density distribution.
Data refers to redshift $z=15.00$ (left), $z=13.01$ (center), and $z=11.00$ (right).
The histograms filled by $45^o$ blue lines refer to pristine gas, while the ones by $135^o$ red lines to enriched gas.
Normalization is fixed to 1 for the maximum, in both cases.
}
\label{fig:GasDistribution}
\end{figure*}
Correspondingly, in Fig. \ref{fig:GasDistribution}, we show the gas distribution (histograms) at redshift $z=15.00$ (left panel), $z=13.01$ (central panel), and $z=11.00$ (right panel).
The histograms filled by red $135^o$ lines refer to enriched gas, and the ones with blue $45^o$ lines to pristine gas.
Most of the particles at low densities are pristine, while the peak of the distribution for metal-enriched gas is at high density, where most of the metals are being produced: these ``hybrid'' particles are the ones where the sub-grid modeling of the multi-phase medium and star formation is being applied. As time passes, metal spreading produces a split, with a second peak at lower and intermediate densities.
This second peak represents the isothermal stage before the run-away collapse.
Its location is very meaningful, since it gives us some idea of the environment affected by metal pollution. The reason for the aggregation of enriched particles in this stage of the phase diagram is due to the long cooling time-scales at low density, in comparison to the cooling branch.
Particles that underwent star formation process are spread around from wind feedback.
They are hot ($T\sim10^5-10^6\,\rm K$) and are ejected into lower-density regions, where their cooling capabilities allow the temperature to rapidly drop down to $T\sim 10^4\,\rm K$.
Subsequent cooling is slightly slower (as observable from the larger spread below $\sim 10^4\,\rm K$, in Fig. \ref{fig:PhaseEvolution}), due to the less efficient metal fine-structure transitions, but it still dominates the total cooling function, with respect to the molecular contribution.
The lack of particles in correspondence of high/intermediate densities and high temperatures is due to the feedback effects combined with strong cooling, which allow the gas to easily cool down to $T\lesssim 10^4\rm\,K$.
Indeed, due to feedback from SN, entropy is injected into the star forming medium and the gas is pushed away towards low-density areas.
The typical velocity for stellar winds is $500\,\rm km/s$, thus in $\sim 10^6-10^7\,\rm yr$ the gas can be pushed several kpc away\footnote{
The physical distance travelled at $500\,\rm km/s$ in $10^6-10^7\,\rm yr$ is $\sim~0.5-5\,\rm~kpc$, corresponding to $\sim~8-80\,\rm kpc$ comoving, at redshift 15.
}
, beyond the isothermal peak (see Fig. \ref{fig:PhaseEvolution}), where densities are orders of magnitude smaller.\\
By comparing the distribution at different redshifts, it turns out that the mass fraction of enriched material located in over-dense (i.e. selected with over-density $\delta\equiv\rho/\rho_{cr}(z)\gtrsim 10^2$) regions is still more than $\sim 99\%$ at $z\sim 13$, and decreases down to $\sim 87\%$ at $z\sim 11$, i.e. $\Delta t \sim 0.08\,\rm Gyr$ later.
At this redshift, the fraction of new metals, $f_{Z,new}$, represents more than $90\%$, and the corresponding enrichment rate, $\dot\varepsilon_Z \equiv \d f_Z/ \d t \simeq f_{Z,new}/\Delta t $, is $\sim 10^{-8}\rm\,yr^{-1}$.
When looking at the amount of metals moving along the cooling branch\footnote{The cooling branch roughly corresponds to densities higher than $\sim 10^2$ times the critical one and below the star formation threshold of about $ 140\,h^2\,{\rm cm^{-3}} \sim 10^{-22}\,h^2\,\rm g/cm^3$.}
we find that $\sim 1\%$, $\sim 6\%$, and $\sim 4\%$ of the enriched mass  at redshift $z=15.00$, $z=13.01$, and $z=11.00$, respectively, is undergoing run-away collapse.\\
Comparison of the simulations with different $Z_{crit}$ shows that this picture does not change dramatically, and the enriched mass undergoing run-away cooling increases, in all the cases, up to around $5\%-7\%$ at $z=13.01$, and reaches about $ 3\%-6\%$ at $z=11.00$.\\
If we look at the run with a Salpeter-like popIII IMF \cite[see details in][in particular Fig. 7]{Maio2010}, stellar lifetimes are longer, and consequently also the time-delay for metal spreading is longer ( $\sim 10^7\,\rm yr$). So, metal-rich particles undergoing run-away cooling are significantly present only after $z\lesssim 14$ and account for $\sim 3\%-8\%$ of the enriched mass.
\\
\begin{figure*}
\centering
\includegraphics[width=0.95\textwidth]{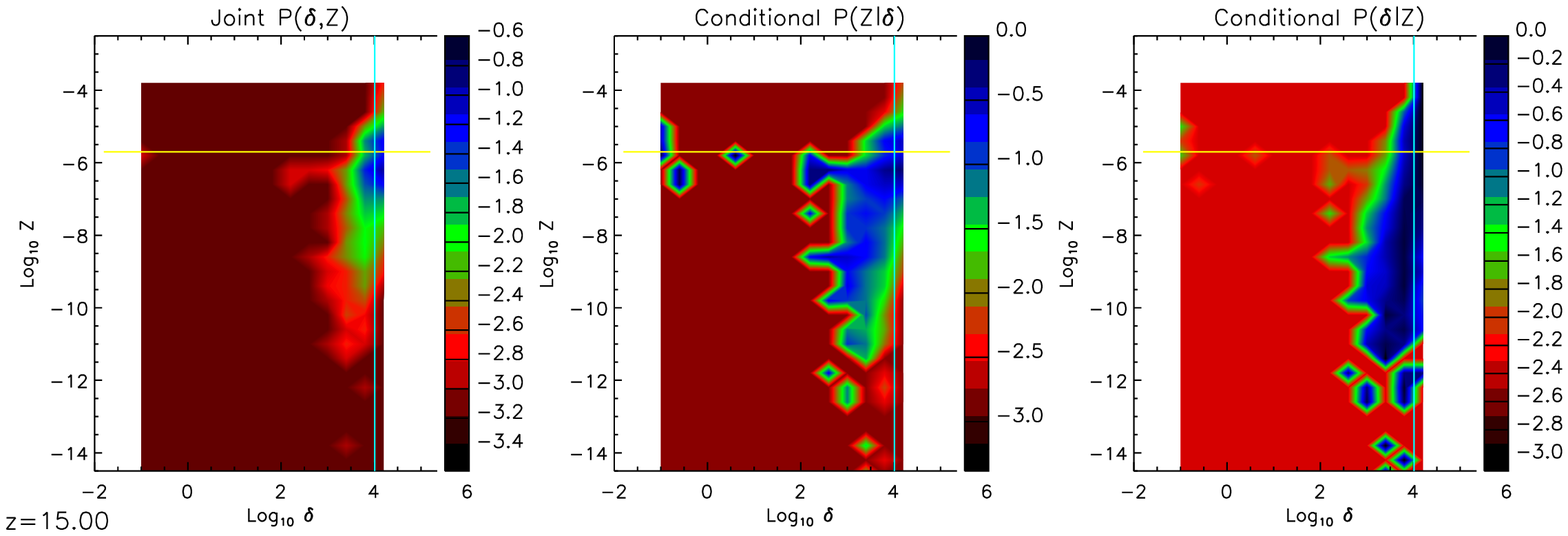}
\includegraphics[width=0.95\textwidth]{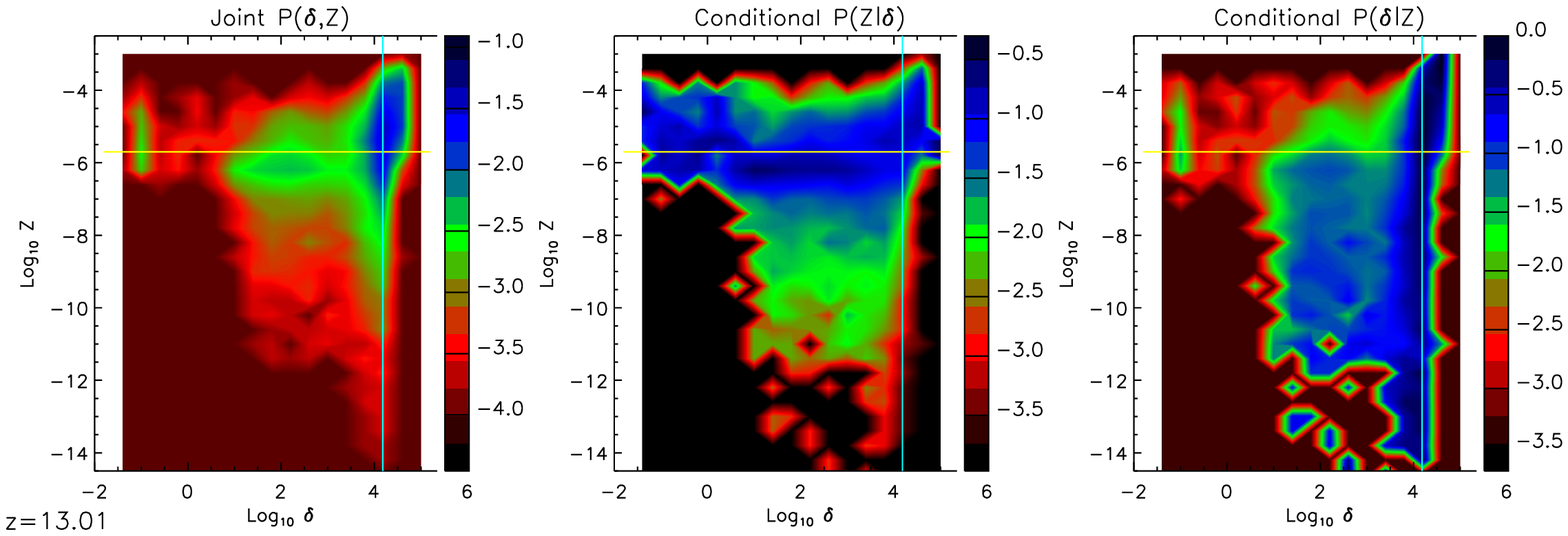}
\includegraphics[width=0.95\textwidth]{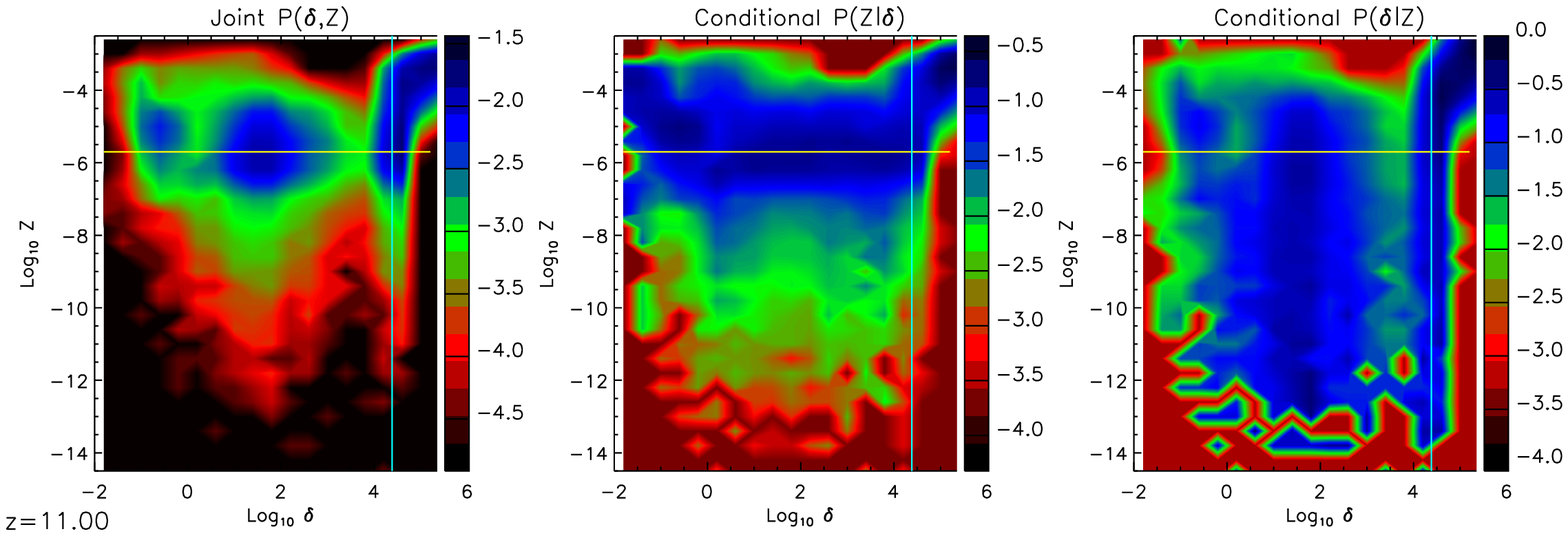}
\caption[Probabilities]{\small
Normalized joint probability distribution (left column), $\rm P(\delta, Z)$, between metallicity, $Z$, and over-density, $\delta$, at redshift $z=15.00$ (top), $z=13.01$ (center), and $z=11.00$ (bottom).
The corresponding conditional probability distributions, $\rm P(Z|\delta)$ and $\rm P(\delta|Z)$, are also shown in the central and right column, respectively.
In each panel, the horizontal, solid line is drawn in correspondence of the critical metallicity, $Z_{crit}=10^{-4}Z_\odot$ and the vertical, solid line marks the star formation over-density threshold.
}
\label{fig:probabilities}
\end{figure*}
In Fig. \ref{fig:probabilities}, we investigate how gas mass metallicity correlates with over-density.
We consider metal-enriched particles and compute the joint probability distribution between $Z$ and $\delta$, $P(\delta,Z)$, and the conditional probability distributions, $P(\delta|Z)$, and $P(Z|\delta)$.
The joint probability distribution, $P(\delta,Z)$, is computed by summing up all the metal contributions at any $\delta$ and $Z$ and normalized over the whole $\{\delta,Z\}$ space.
While the conditional probability $P(Z|\delta)$ is computed by summing up all the contributions for a given $\delta$, and normalized over the corresponding $\{Z\}$ subspace.
Similarly, the conditional probability $P(\delta|Z)$ is computed by summing up all the contributions for a given $Z$, and normalized over the corresponding $\{\delta\}$ subspace.
The horizontal line marks the critical value, $Z_{crit}$, while the vertical line marks the star formation density threshold ($\sim 140 h^2 \rm cm^{-3}$), above which the star formation sub-grid modeling is applied.
The plots refer to redshift $z=11$ (upper row), $z=13$ (central row), and $z=15$ (bottom row).
The joint probability distribution, $P(\delta,Z)$, has a maximum around $Z\sim 10^{-6}-10^{-4}$, almost despite of the redshift considered, and its time evolution shows clearly how metals are ejected by wind feedback from the star forming regions, at $\delta\sim 10^4-10^5$, to the lower-density ones, at $\delta \lesssim 10^2$.
This process is better highlighted in the conditional probability distribution $P(Z|\delta)$: at earlier times, during the very first episodes of star formation, there are still few metals, mostly at low $Z$, and they have not reached yet the low-density environments. Later on, $P(Z|\delta)$ is rapidly boosted to higher metallicities, with $Z$ overcoming $Z_{crit}$ at any $\delta$, as a consequence of the high metal yields of primordial stars.
By checking the conditional probability $P(\delta|Z)$, it emerges that, for any fixed $Z$ besides the star-forming particles at $\delta\sim 10^4-10^5$, the regions where enrichment due to metal-spreading is dominant are soon the ones with $\delta\lesssim 10^2$ (see also previous sections).
Furthermore, the over-density range between $\sim 10^2$ and $\sim 10^4$, at $z\sim 11$, presents a gap in the distribution, because it corresponds to the fast, catastrophic run-away cooling, through which enriched gas rapidly re-collapses to higher densities and lower temperatures.

\subsubsection{PopIII and popII}
\begin{figure*}
\centering
\includegraphics[width=0.3\textwidth]{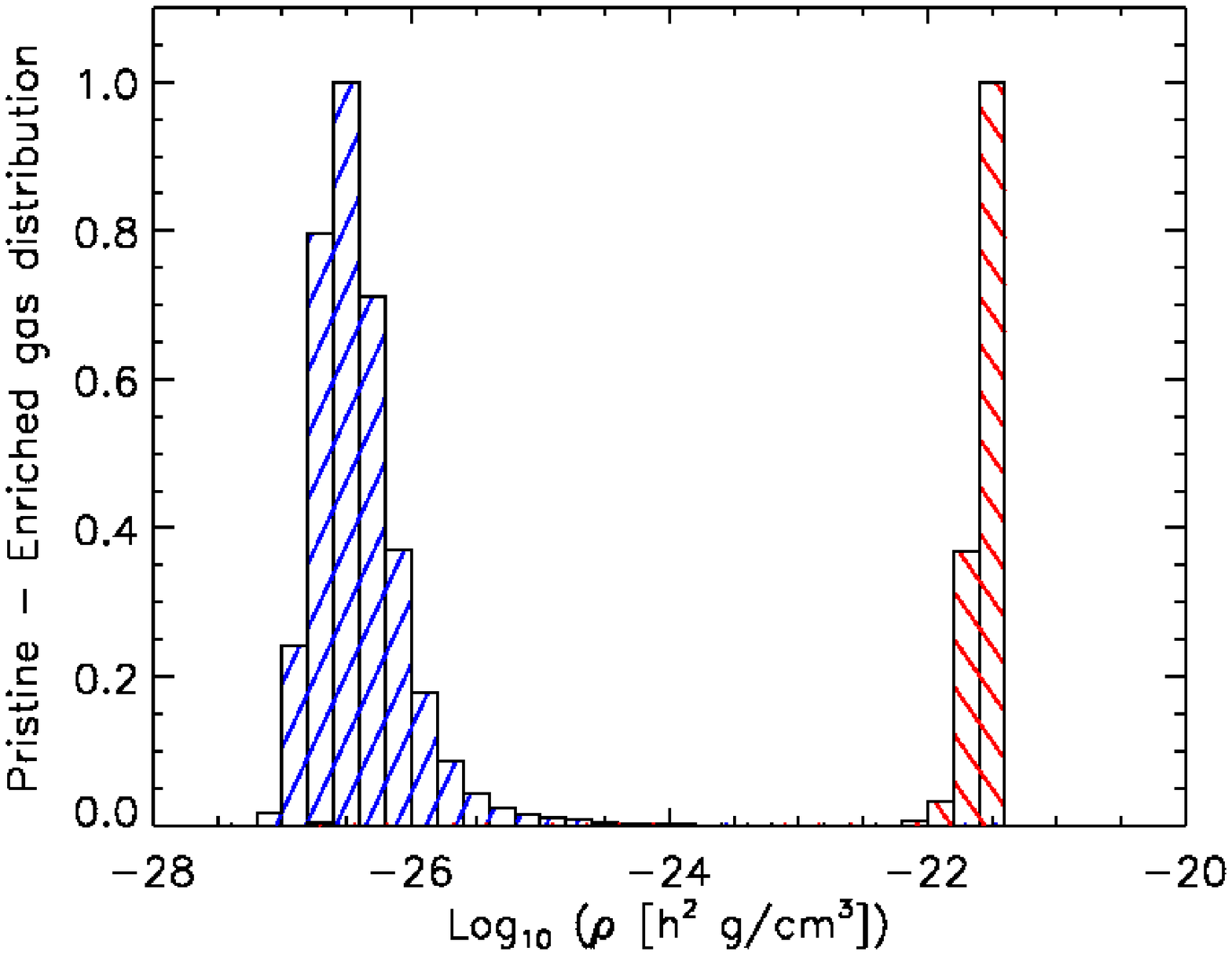}
\includegraphics[width=0.3\textwidth]{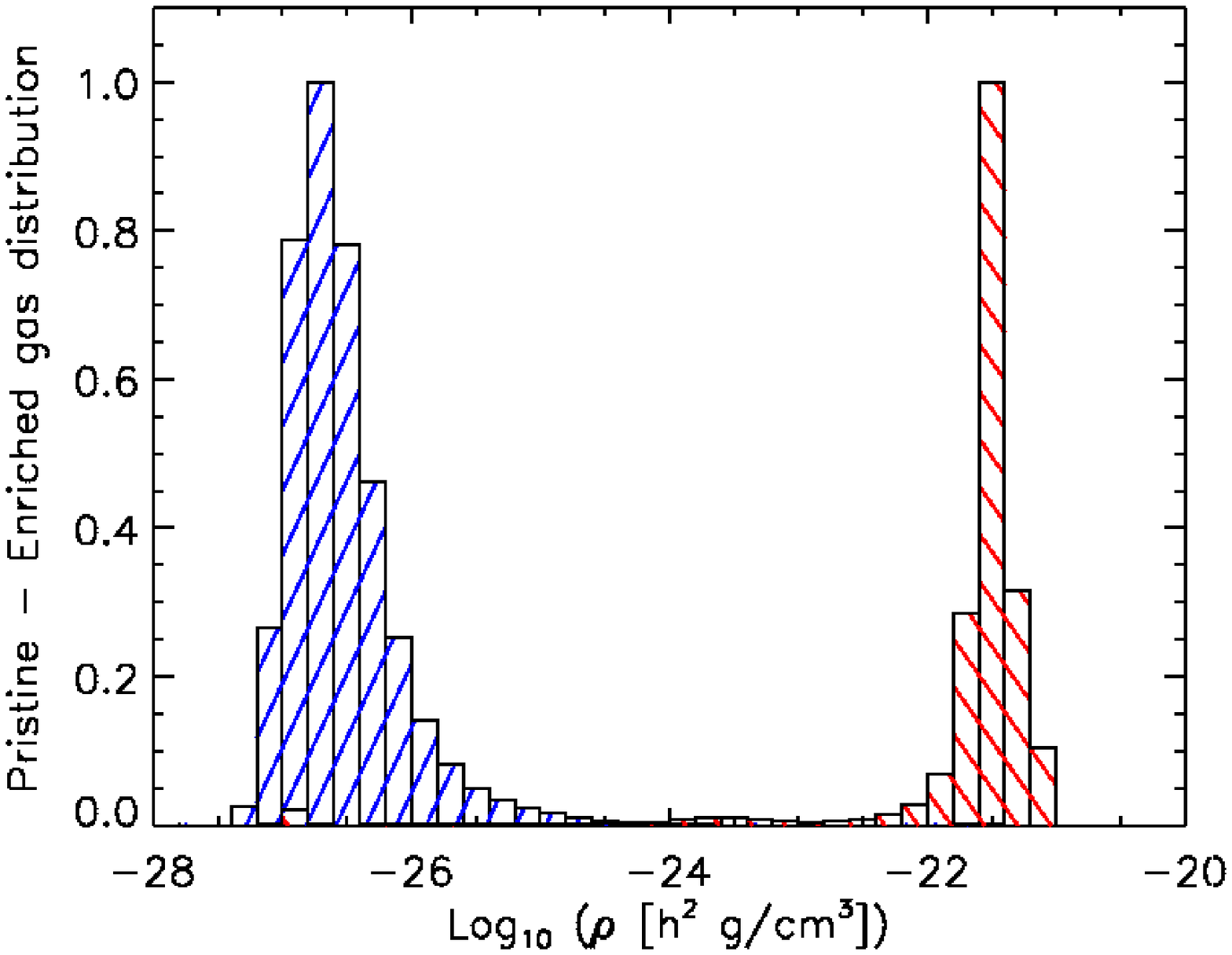}
\includegraphics[width=0.3\textwidth]{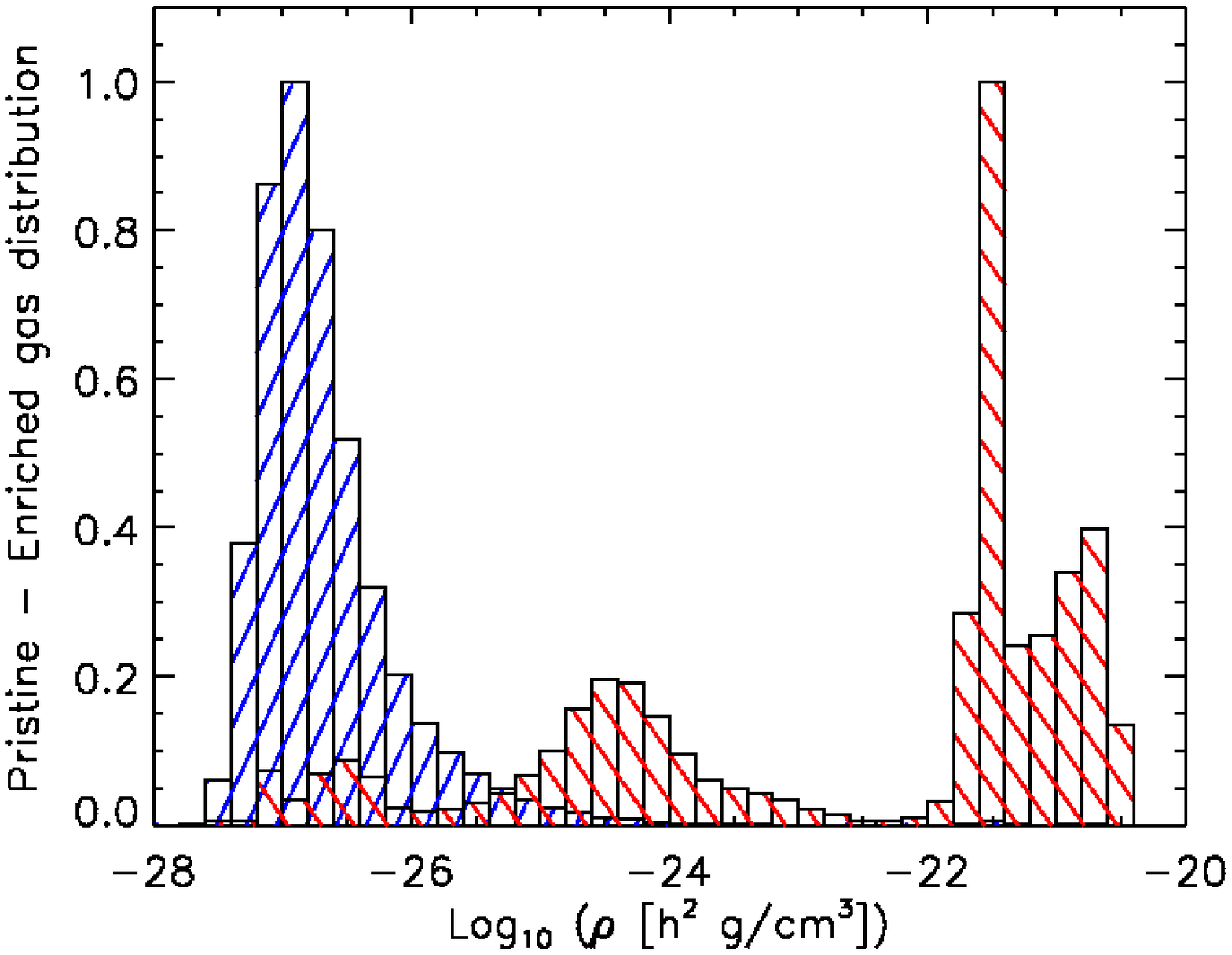}\\
\includegraphics[width=0.3\textwidth]{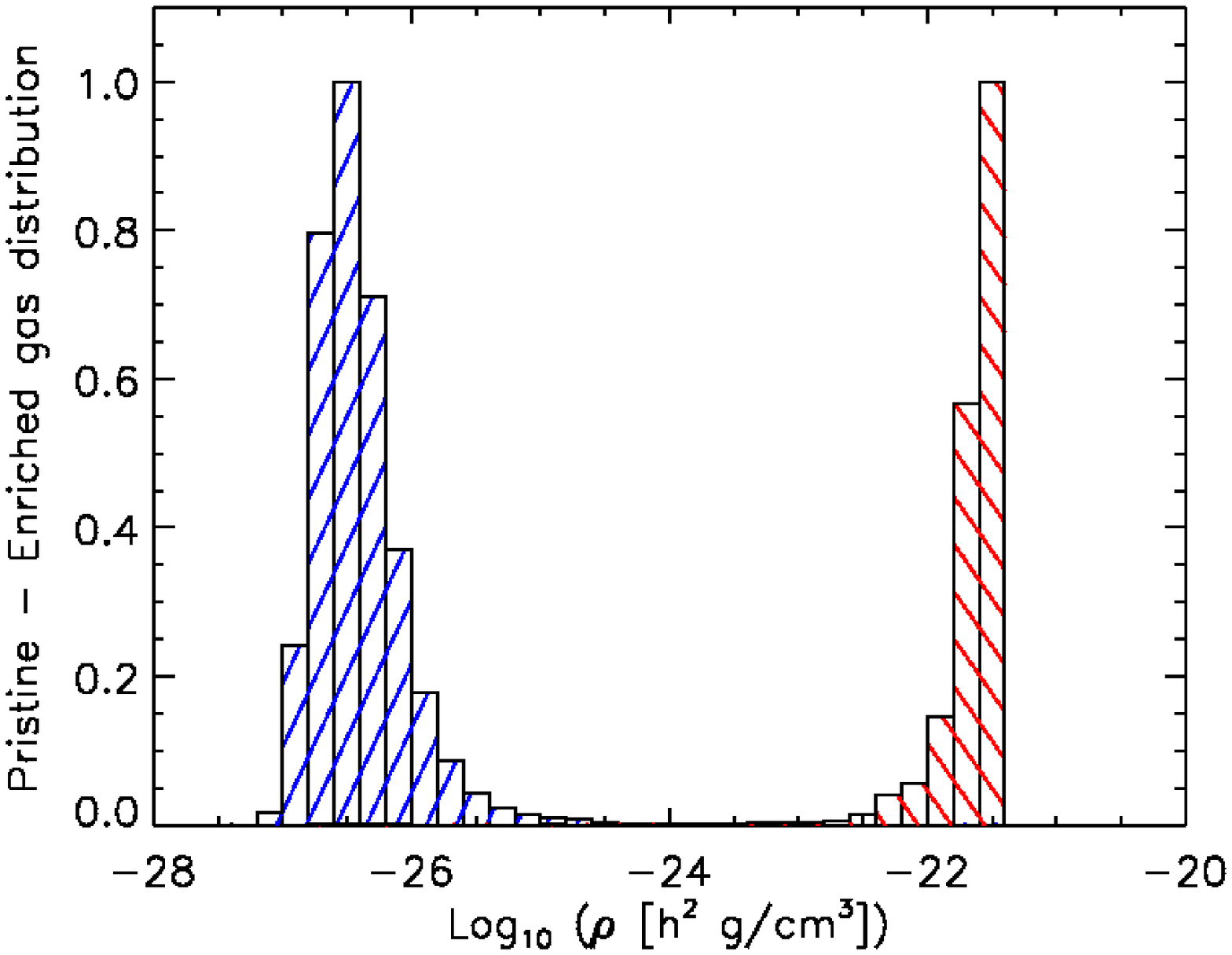}
\includegraphics[width=0.3\textwidth]{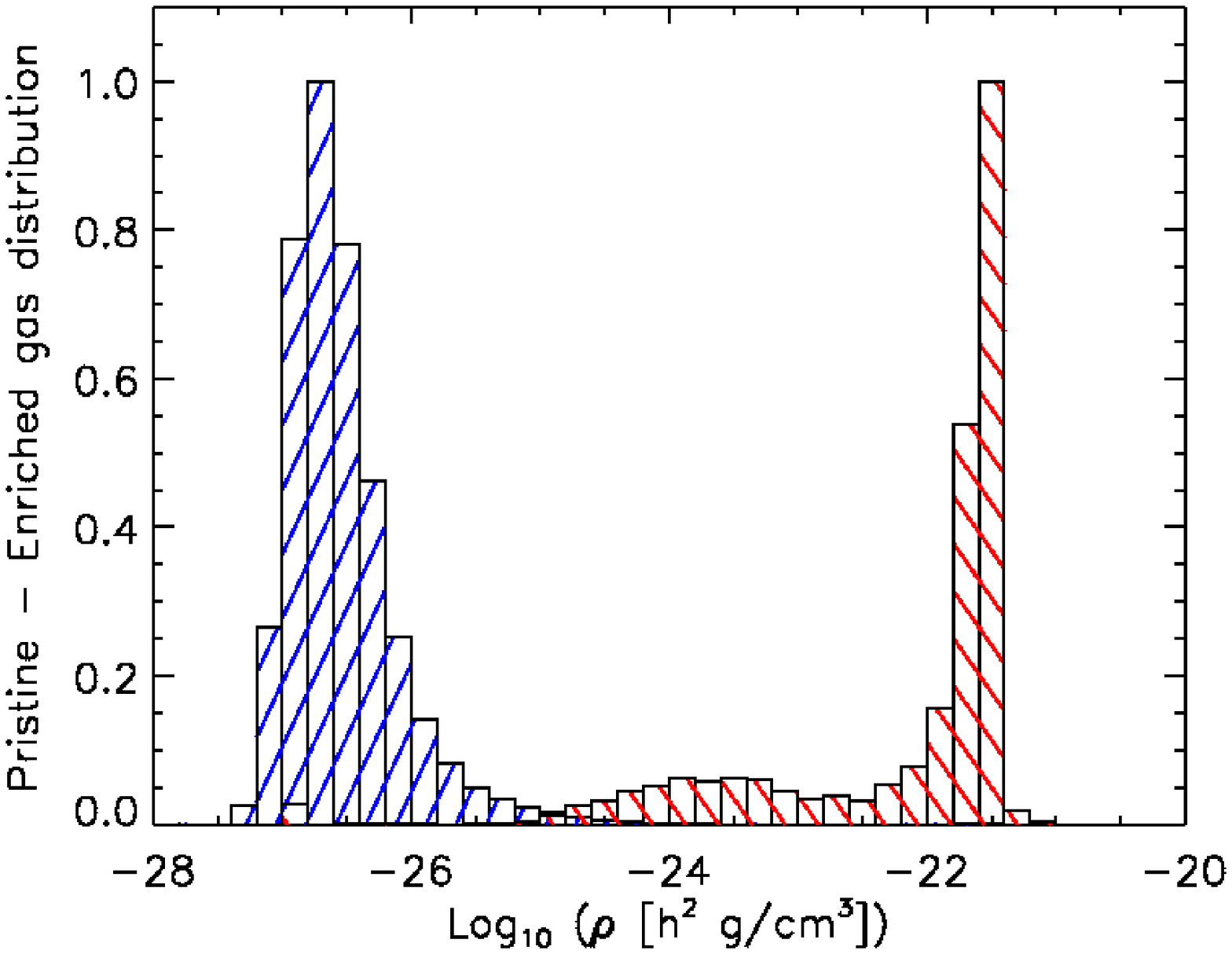}
\includegraphics[width=0.3\textwidth]{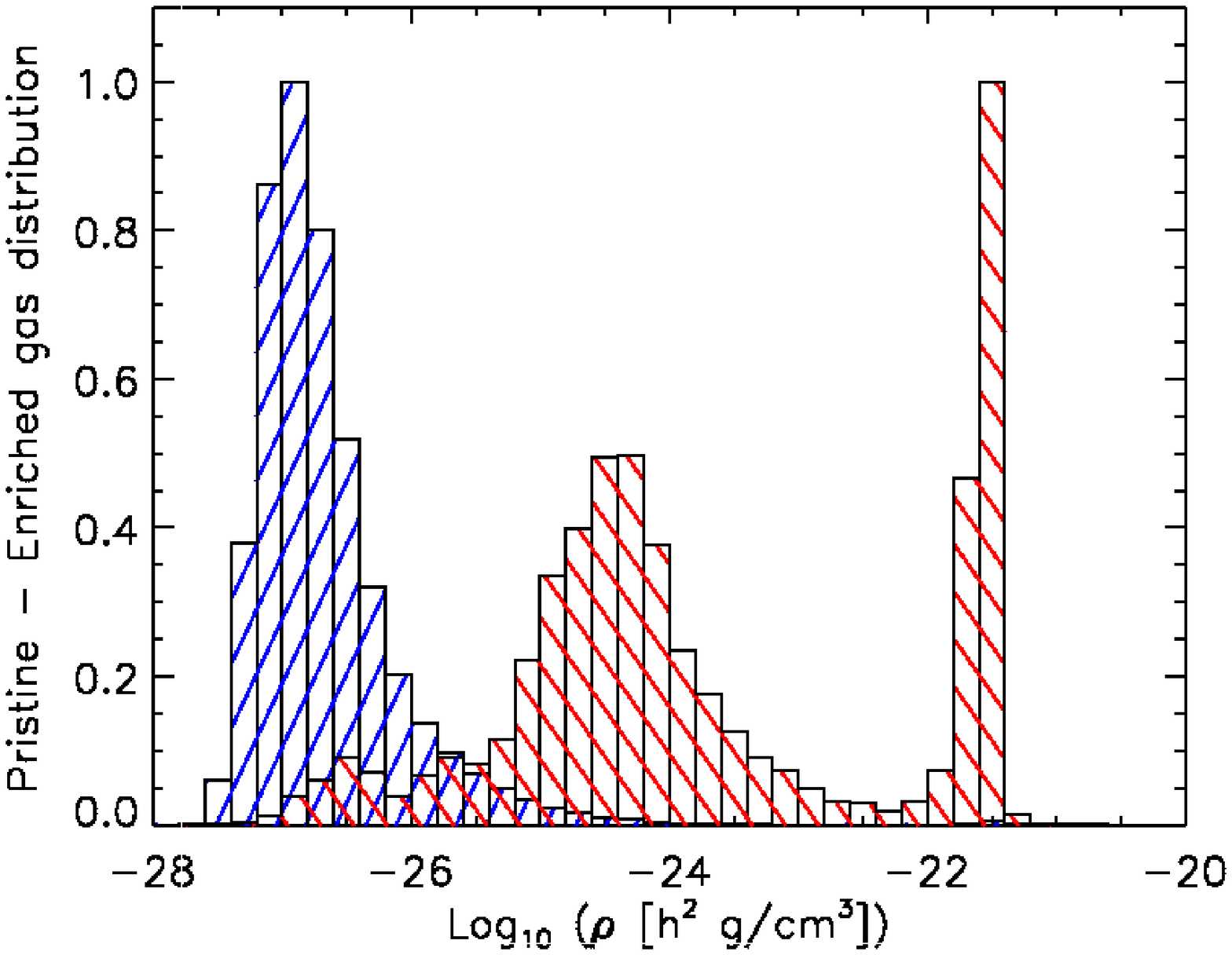}
\caption[Distribution $Z_{crit}$]{\small
Redshift evolution of gas density distributions of pristine and metal-rich particles, selected by their $Z$.
In both rows, we show the pristine-gas distribution with histograms filled by $45^o$ blue lines, while enriched gas distributions, for particles having $Z \ge Z_{crit}$ in the upper row and $Z<Z_{crit}$ in the lower row, are plotted with hystograms filled by $135^o$ red lines.
Data refer to  $z=15.00$ (left), $z=13.01$ (center), and $z=11.00$ (right).
Normalization is fixed to 1 for the maximum.
}
\label{fig:GasDistributionZvsZcrit}
\end{figure*}
After our investigation of the global features of the primordial enrichment, we want to study the differences between population III and population II regimes.\\
In Fig. \ref{fig:GasDistributionZvsZcrit}, we select particles according to their metallicity and plot the density distribution of metal-rich particles for both popII and popIII ones (in a similar way to Fig. \ref{fig:GasDistribution}).
By comparing the distributions of particles that have $Z\ge Z_{crit}$ (first row)  and $Z<Z_{crit}$ (second row), it is easily shown that, in both cases, enriched star forming particles are continuously produced by the ongoing star formation, and, while the Universe evolves, wind feedback pushes them towards external, lower density regions \cite[see][]{Tornatore2007}.
In particular, the relative fraction of popIII particles ouside the collapsed objects is more than a factor of 2 higher than the popII one.
Let us discuss these two cases more in detail (see also Table \ref{tab:fractions}).
\begin{table}
\centering
\caption[Fractions]{Mass fractions of enriched gas.}
\begin{tabular}{lcc}
\hline
\hline
Redshift & popII  		& popIII   \\
	 & relative fraction 	& relative fraction \\
\hline
15 & 43\% &  57\% \\	
13 & 94\% &   6\% \\	
11 & 99\% &  1\% \\	
\hline
\label{tab:fractions}
\end{tabular}
\begin{flushleft}
\vspace{-0.5cm}
{\small
}
\end{flushleft}
\end{table}

\begin{itemize}
\item
{\it PopII}.
The mass fraction of metals with $Z\ge Z_{crit}$ is initially zero, but reaches already $\sim 43\%$ at $z\sim 15$ and rapidly rises up to $\sim 94\%-99\%$ at $z\sim 13-11$, and above at later times.
The polluted material is pushed away from the over-dense regions 
and expands with time.
In particular, about $\sim 23\%$ of formed popII material has travelled into surrounding $\delta<10^2$ regions, by redshift $z\sim 11$ -- a time lag of $\sim 2\times 10^8\,\rm yr$.
\item
{\it PopIII}.
For the very-low metallicity particles, the picture is quantitatively quite different and particles with $Z<Z_{crit}$ are found outside the collapsing regions, in a more significant way.
The onset of star formation is initially dominated by popIII, but shortly after, at $z\sim 15$, the residual enriched mass fraction with $Z<~Z_{crit}$ drops down $\sim 57\%$, and becomes just $\sim 6\%$ at $z\sim 13$, and  $\sim 1\%$ at $z\sim 11$.
\end{itemize}

\begin{table}
\centering
\caption[Fractions]{Mass fractions of the enriched cooling gas.}
\begin{tabular}{lccc}
\hline
\hline
Redshift 	& popII cooling 	& popIII cooling\\
 		& relative fraction 	& relative fraction\\
\hline
15 & 3\%  &  97\% \\
13 & 80\% &  20\% \\
11 & 94\% &   6\% \\
\hline
\label{tab:cooling}
\end{tabular}
\begin{flushleft}
\vspace{-0.5cm}
{\small
}
\end{flushleft}
\end{table}
Let us now focus on the properties of the enriched cooling mass only (see Table \ref{tab:cooling}), i.e. the mass which has experienced wind feedback and is cooling back into star forming sites.
In the initial stages, as mentioned in Sect. \ref{sect:statistics}, roughly $\sim 1\%$ of the enriched mass is in the catastrophic cooling regime, and this fraction slightly increases up to $4\%-6\%$ at $z\sim 11$.
At $z\sim 15$, the enriched cooling mass is made of popIII type for $\sim 97\%$ and of popII for $\sim 3\%$.
After that, the situation completely changes and, at $z\sim 13$, 
the popIII part is only $\sim 20\%$, and the popII rapidly increases up to $\sim 80\%$.
Later on, at $z\sim 11$, the residual popIII mass fraction drops down to $\sim 6\%$ and the popII one increases up to $\sim 94\%$.
At lower redshifts, the popIII part becomes negligeble with respect to the popII one.
Also in this case, different values for $Z_{crit}$ do not induce drastic changes: the popIII material is significantly present only at the very first stages and completely replaced by popII at $z\lesssim 14-13$, although occurs more quickly for lower $Z_{crit}$.
\\
When a Salpeter-like popIII IMF is adopted, the residual collapsing popIII material at $z\lesssim 14$ is again only a few percent (more precisely, $\sim 3\%$ at $z\sim 11$).


\subsection{Dynamics}\label{sect:dynamics}

\begin{figure*}
\centering
\includegraphics[width=\textwidth]{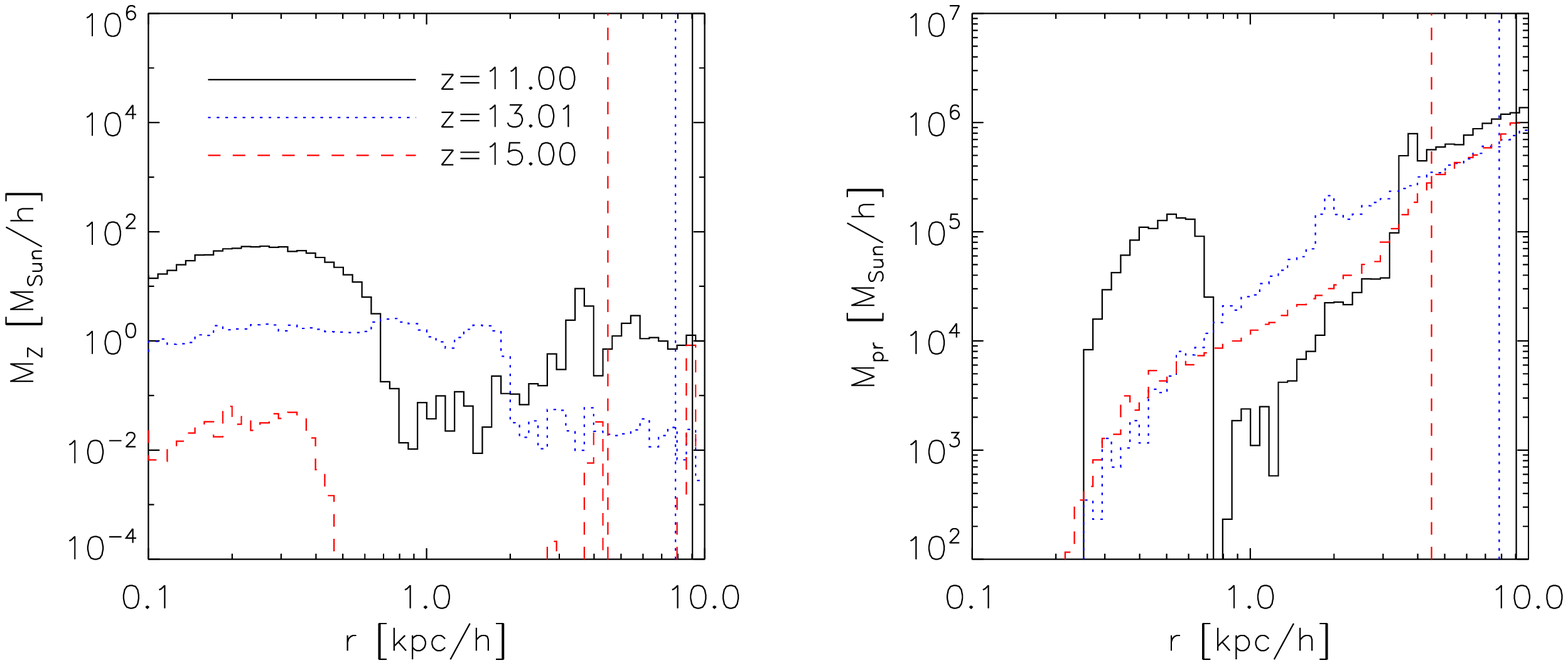}
\caption[Shells]{\small
Mass profiles as a function of comoving radius for enriched (left panel) and pristine (right panel) gas at redshift $z=11.00$ (solid lines), $z=13.01$ (dotted lines), and $z=15.00$ (dashed lines).
Each vertical line corresponds to the virial radius at each redshift.
The comoving gas softening length is $\sim 0.1\,\rm kpc/{\it h}$ and the minimum gas smoothing length is $\sim 0.14\,\rm kpc/{\it h}$.
}
\label{fig:shells}
\end{figure*}
\begin{figure*}
\centering
\includegraphics[width=0.33\textwidth]{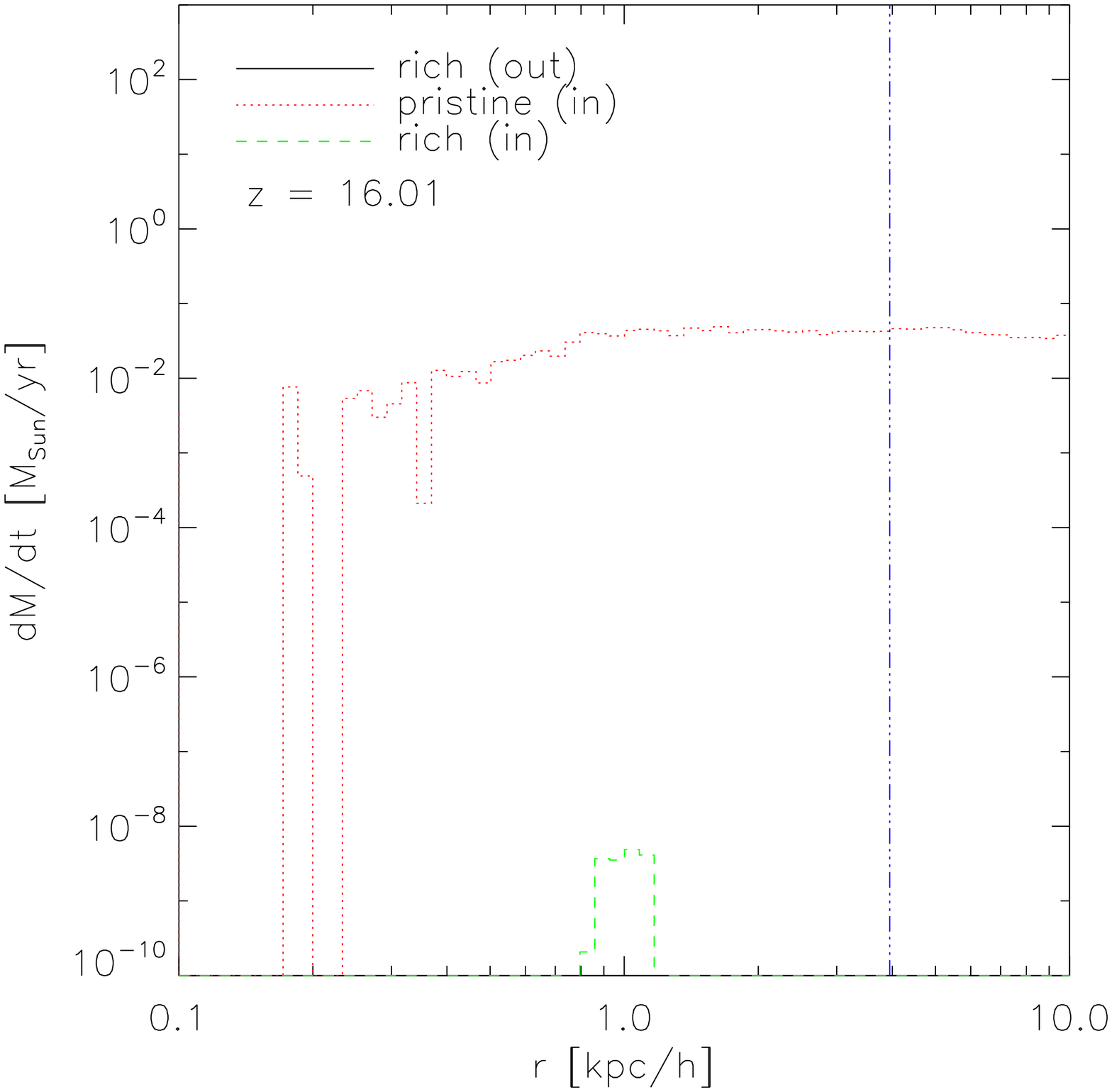}
\includegraphics[width=0.33\textwidth]{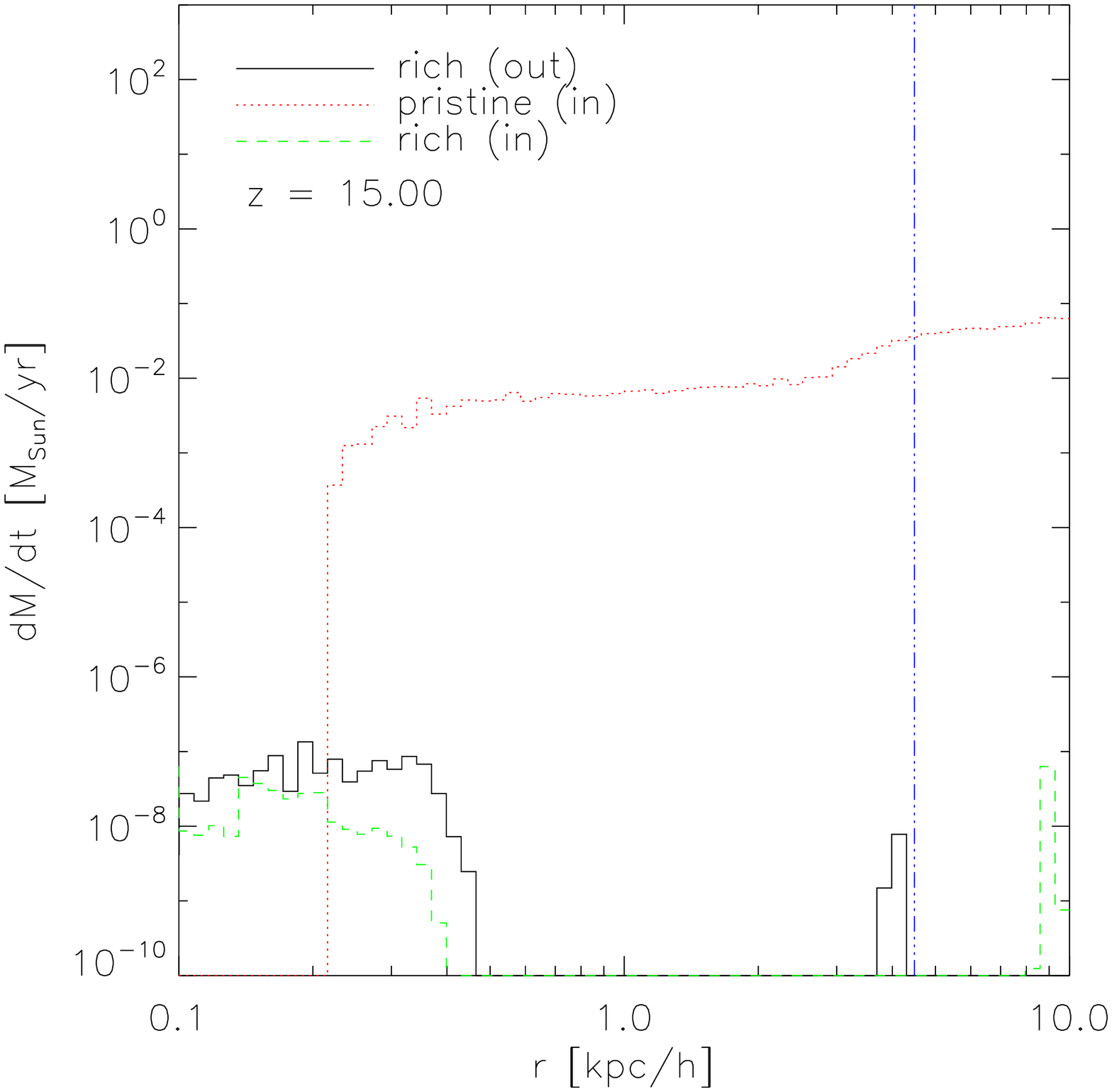}
\includegraphics[width=0.33\textwidth]{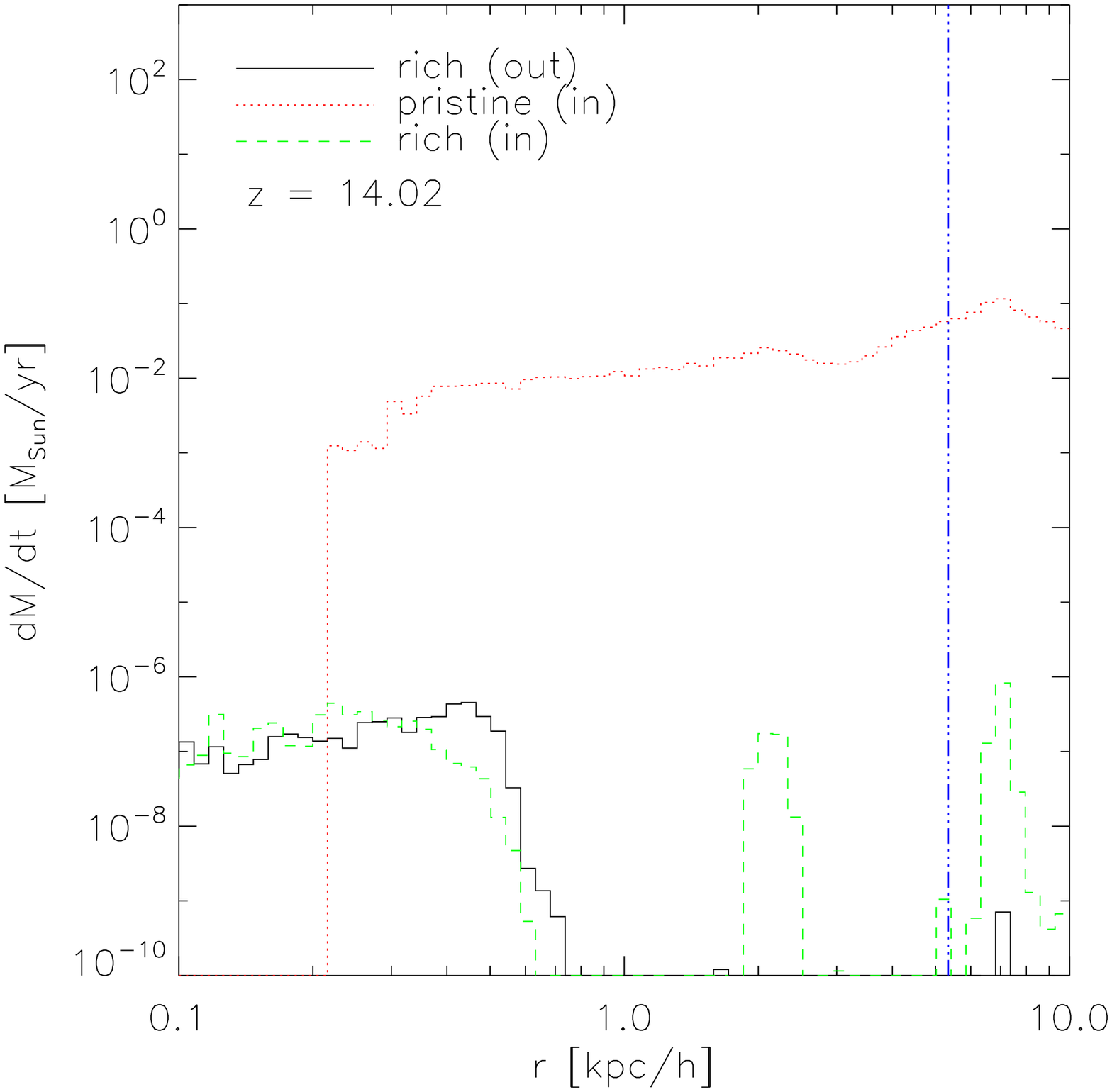}
\includegraphics[width=0.33\textwidth]{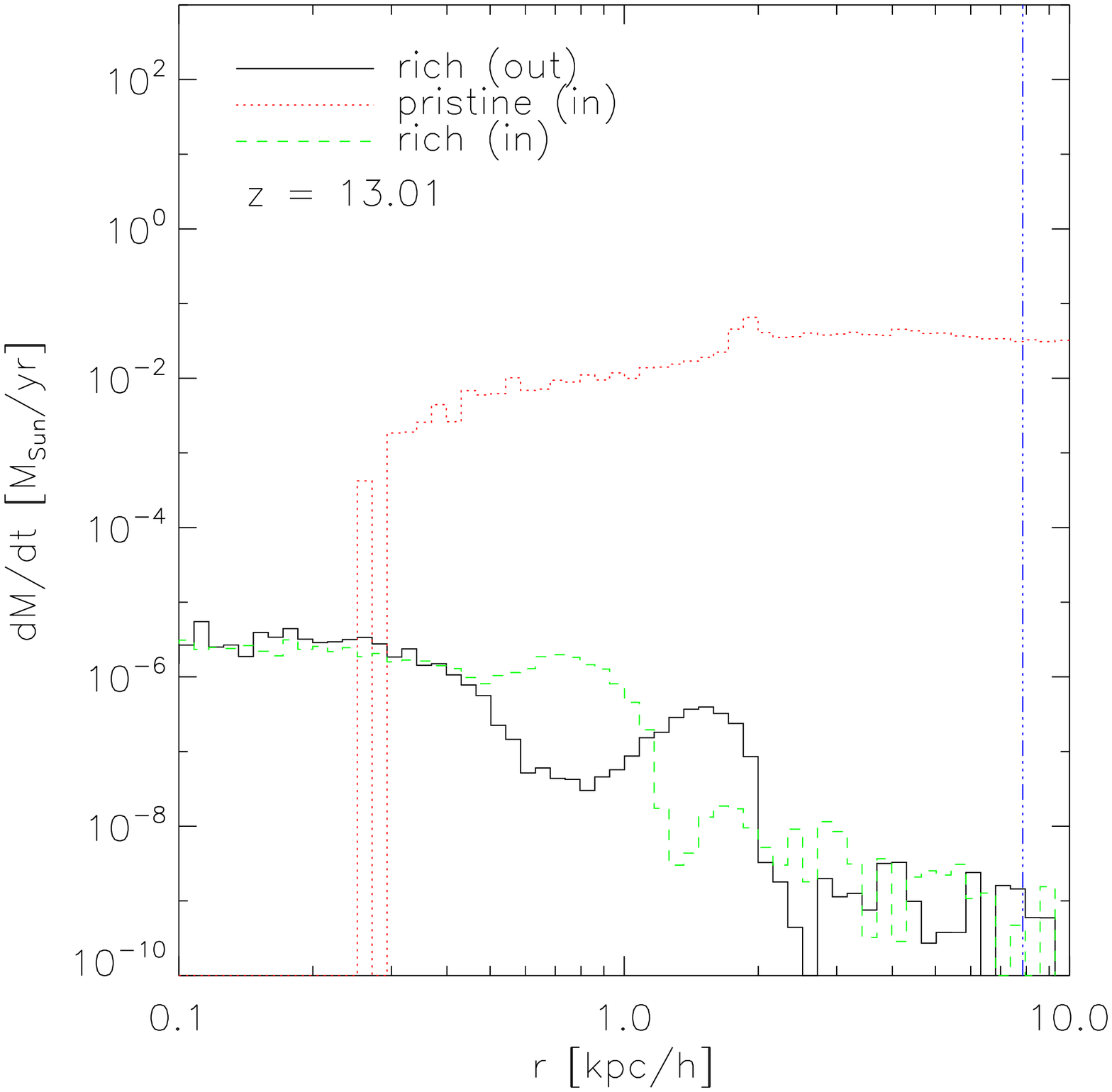}
\includegraphics[width=0.33\textwidth]{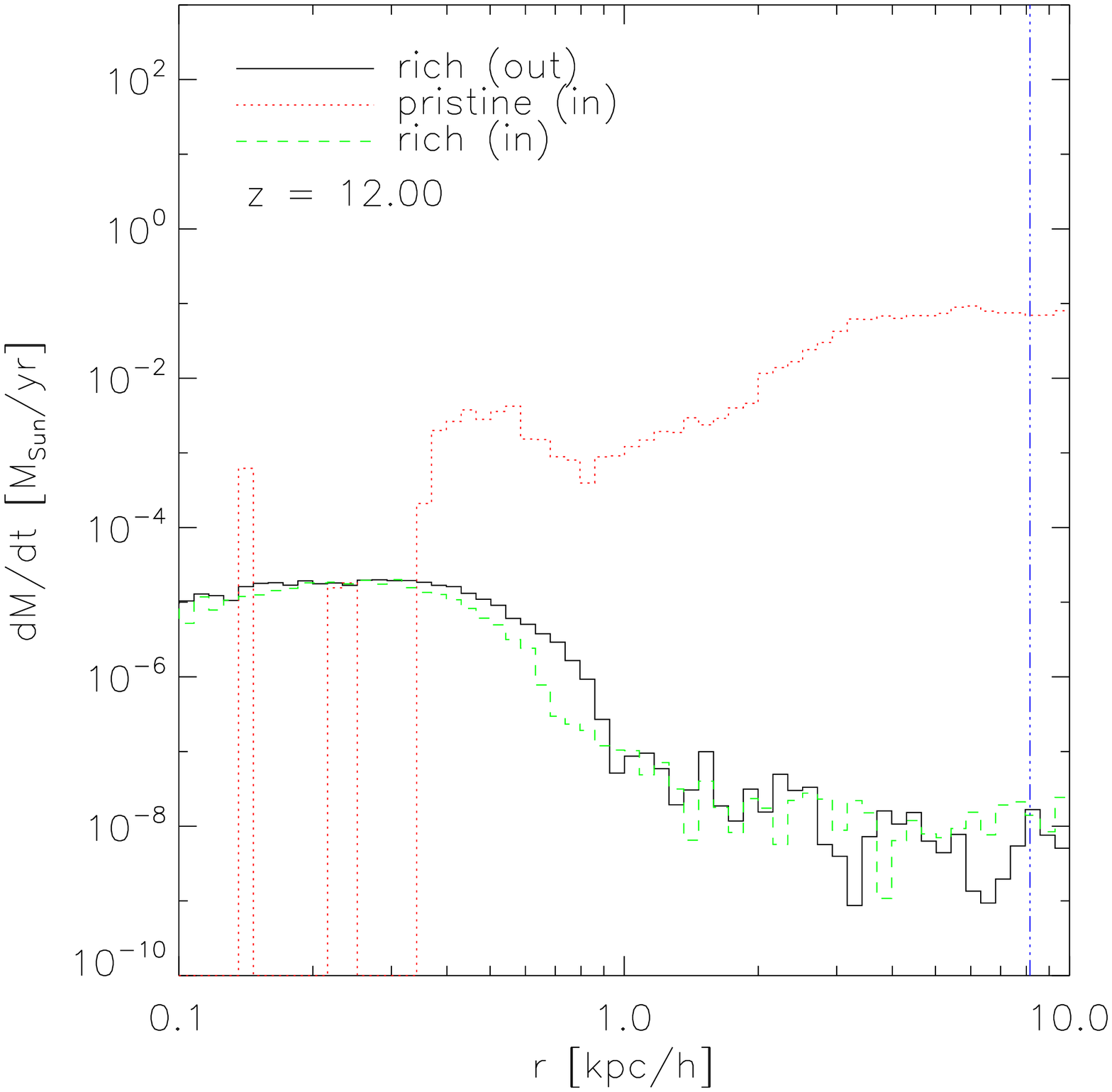}
\includegraphics[width=0.33\textwidth]{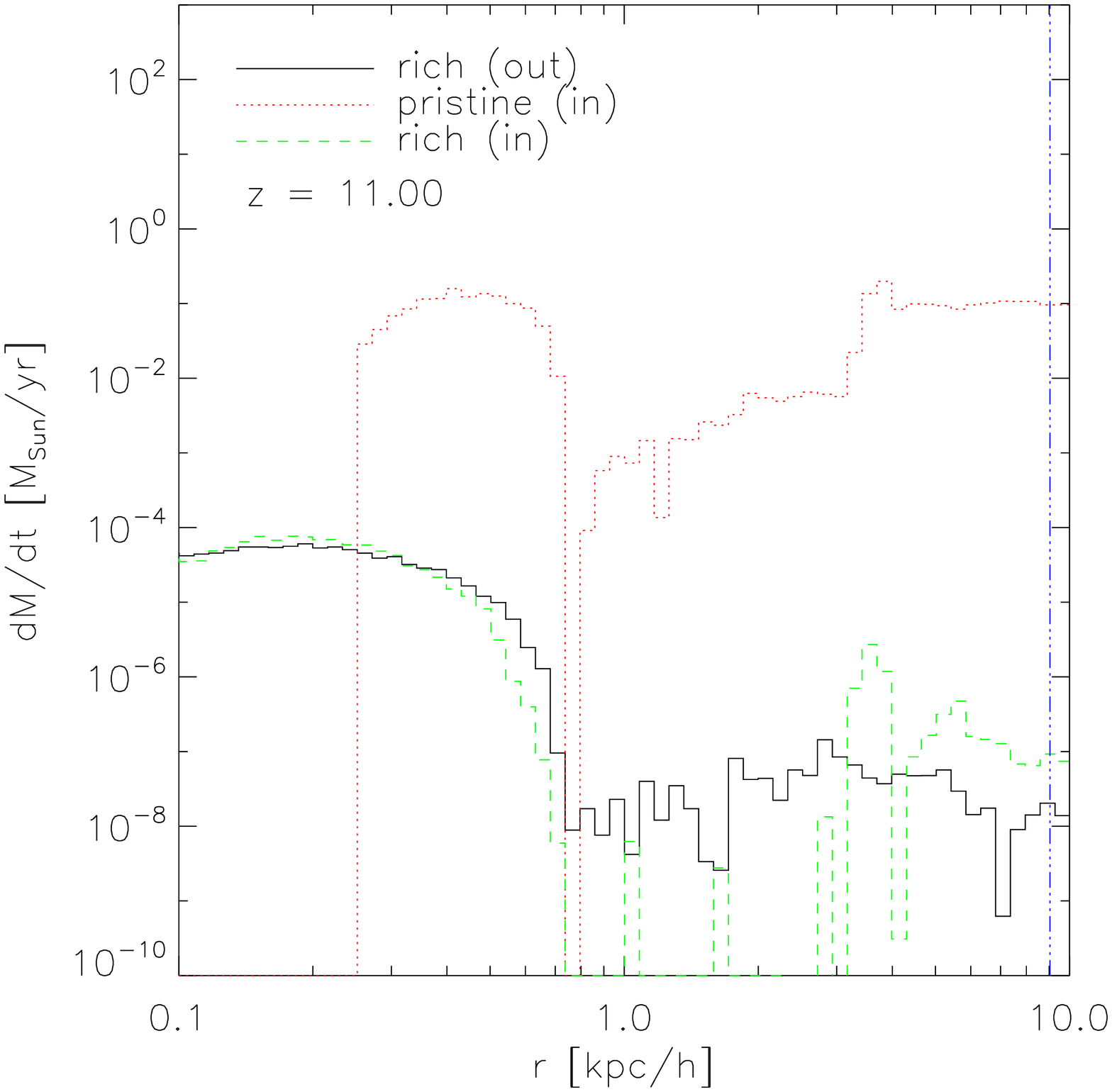}
\caption[Shells]{\small
Mass outflow rates for enriched gas (solid lines), and inflow rates for both pristine gas (dotted lines)  and enriched gas (dashed lines).
Redshifts are
$z=16.01$ (upper left), $z=15.00$ (upper center), $z=14.02$ (upper right),
$z=13.010$ (lower left), $z=12.00$ (lower center), and $z=11.00$ (lower right).
The vertical dot-dashed lines are the corresponding virial radii.
}
\label{fig:rates}
\end{figure*}
\begin{figure*}
\centering
\includegraphics[width=0.29\textwidth]{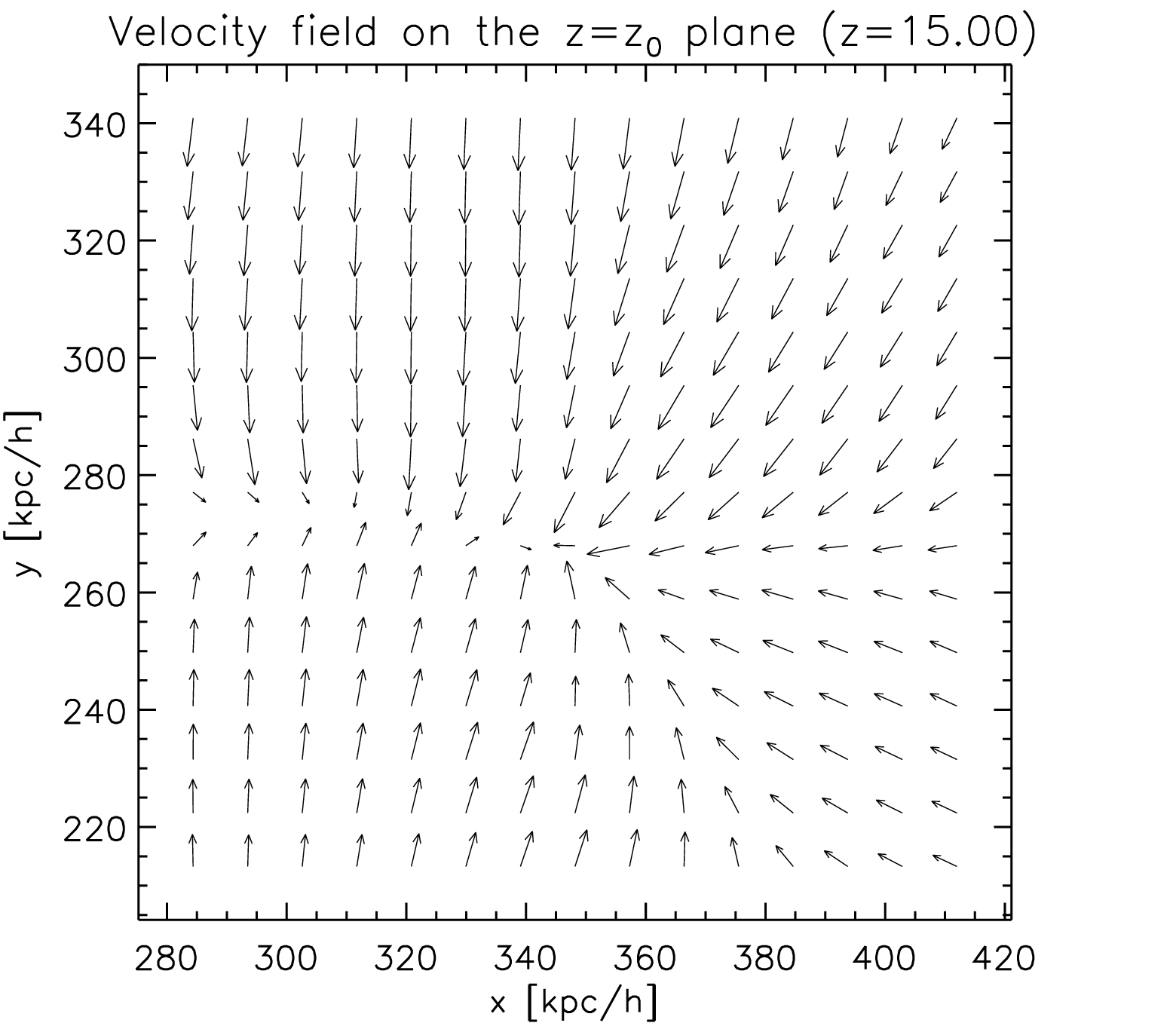}
\includegraphics[width=0.29\textwidth]{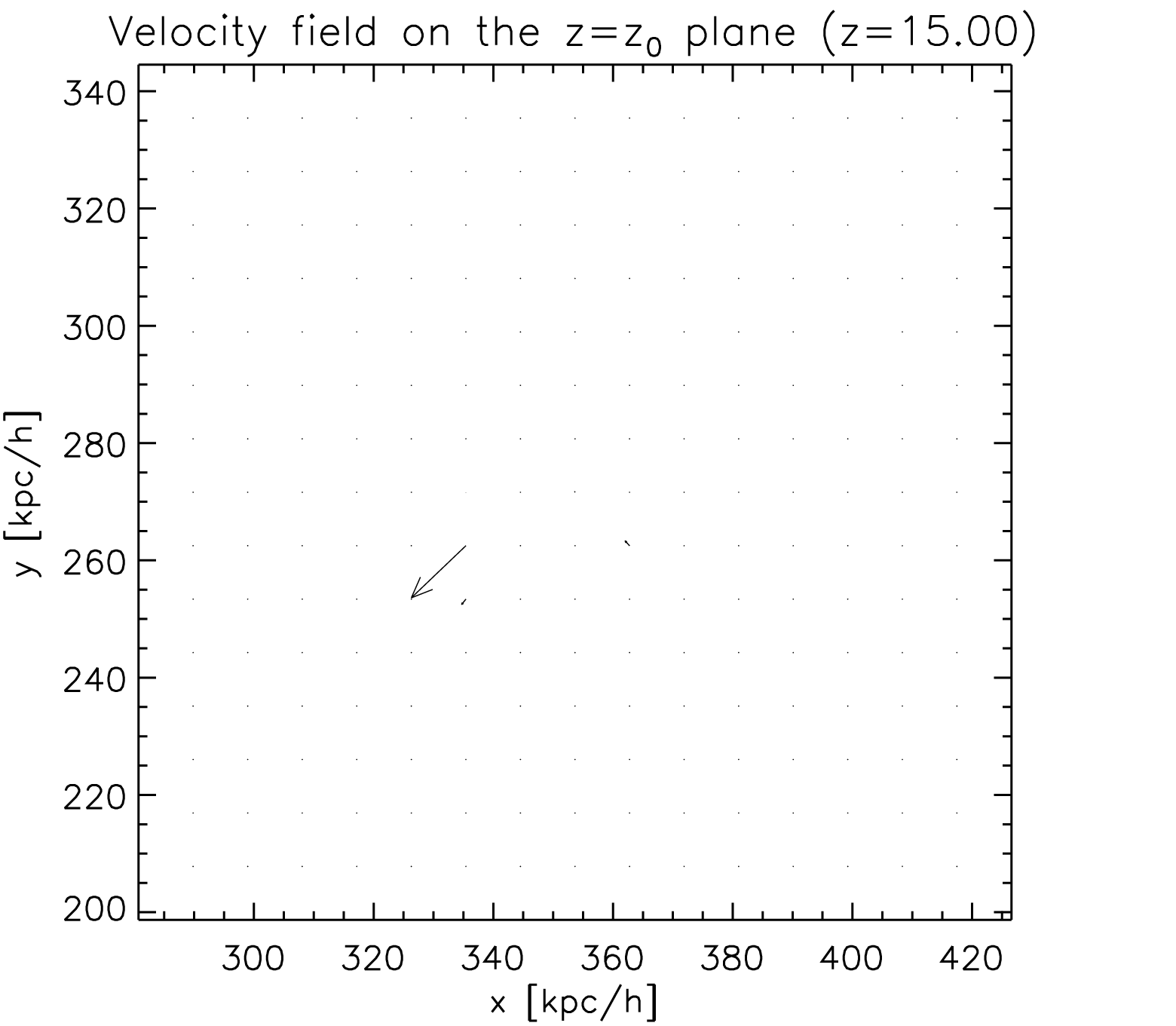}
\includegraphics[width=0.29\textwidth]{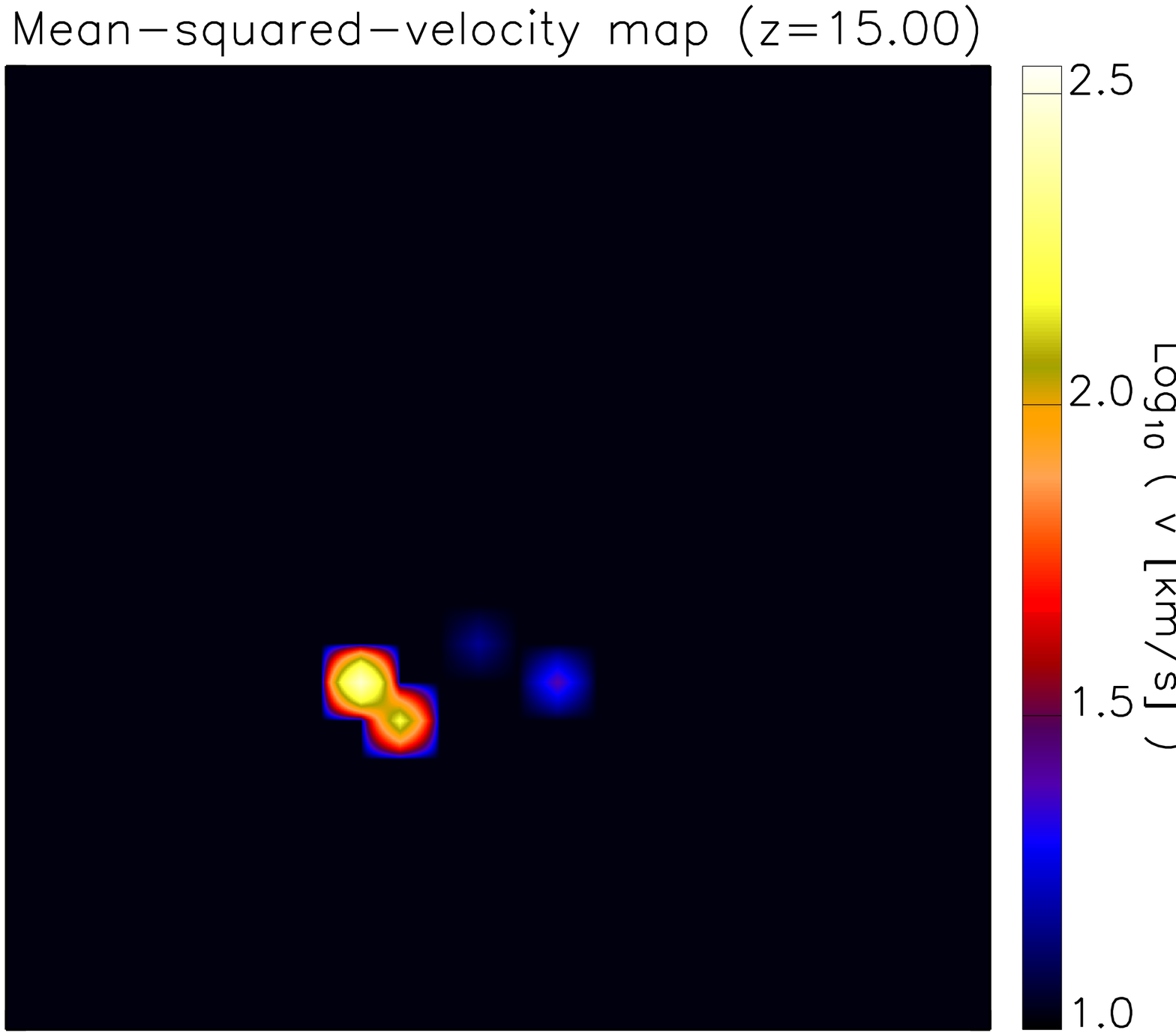}
\hspace{1cm}\\
\includegraphics[width=0.29\textwidth]{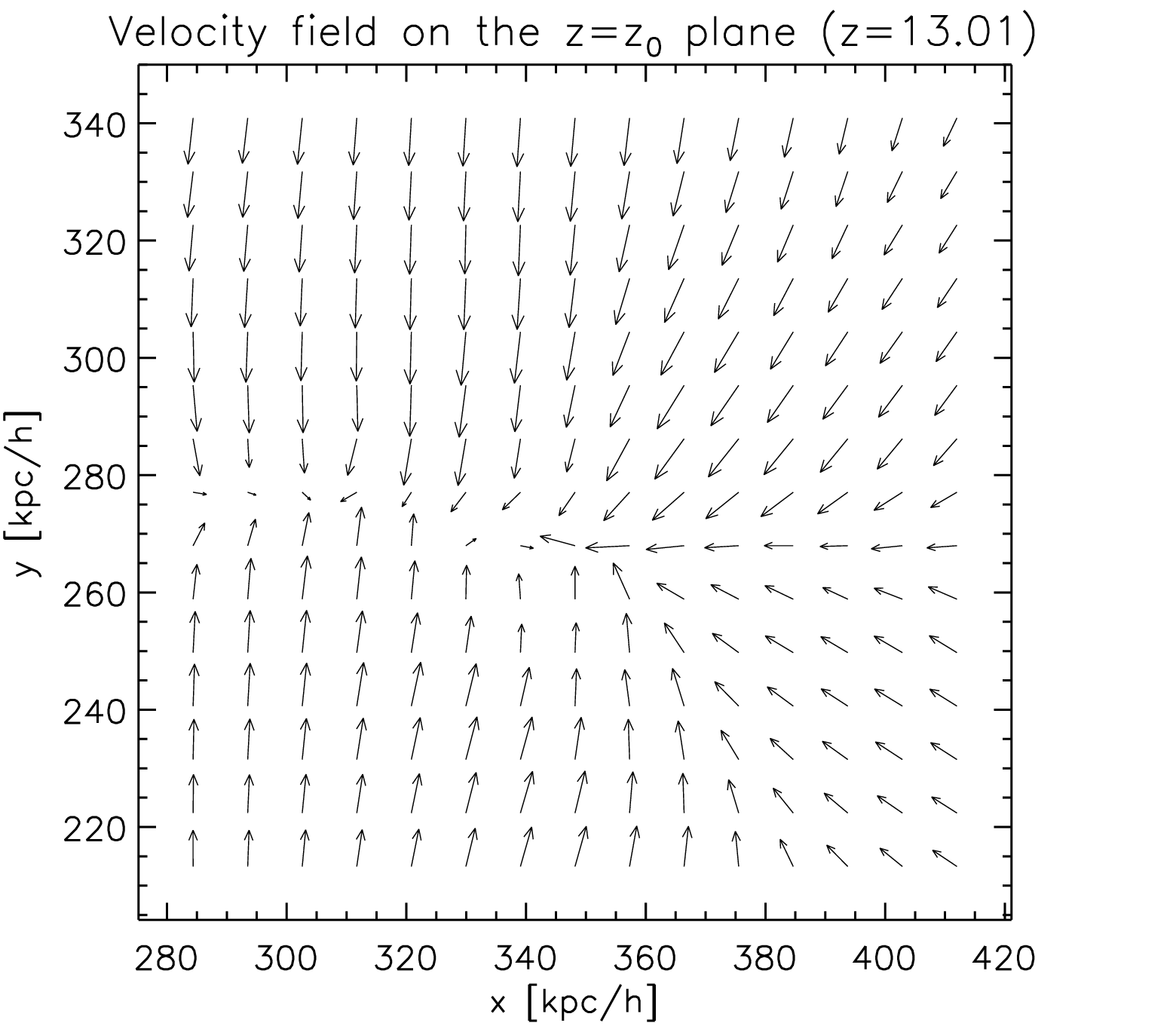}
\includegraphics[width=0.29\textwidth]{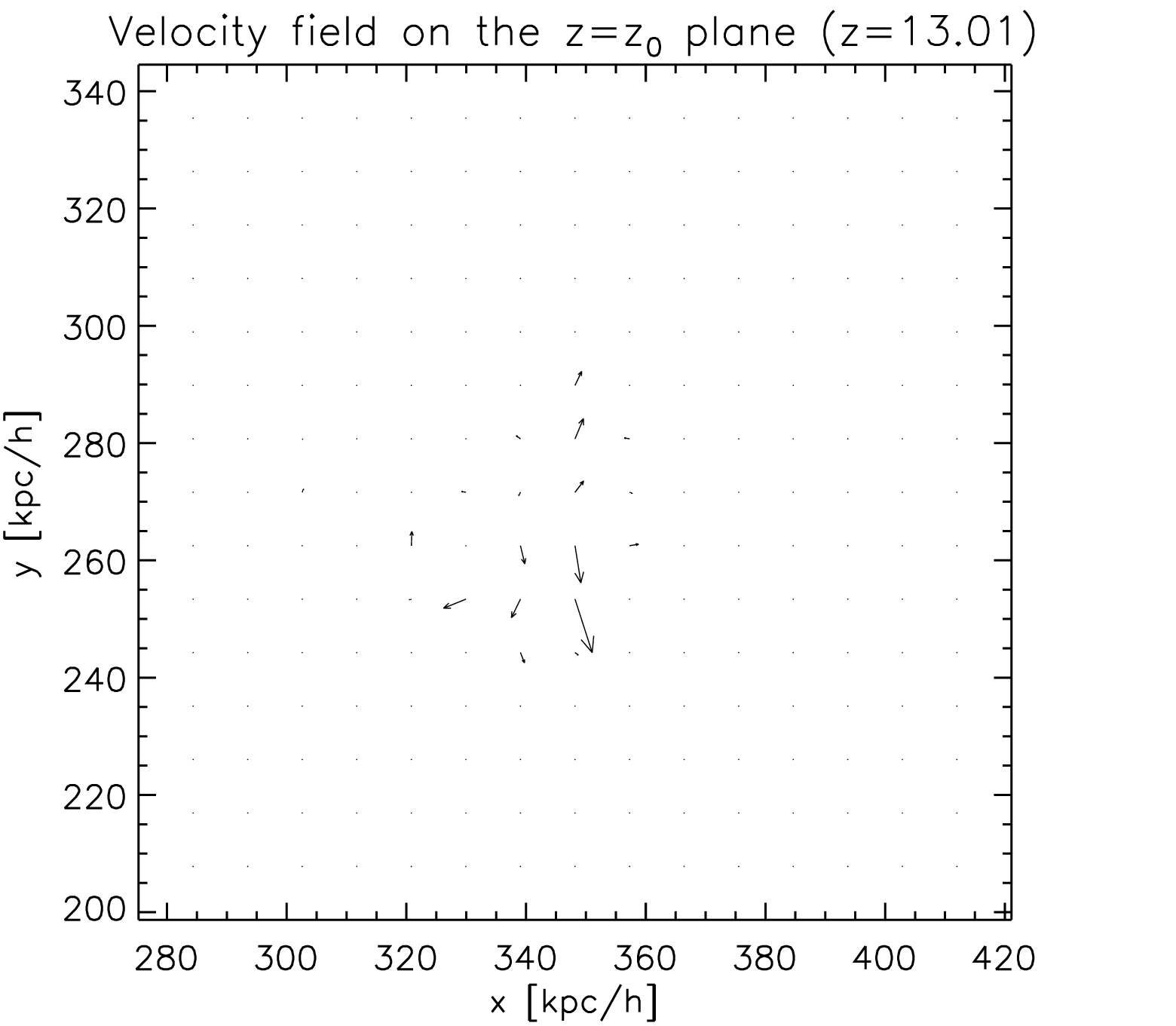}
\includegraphics[width=0.29\textwidth]{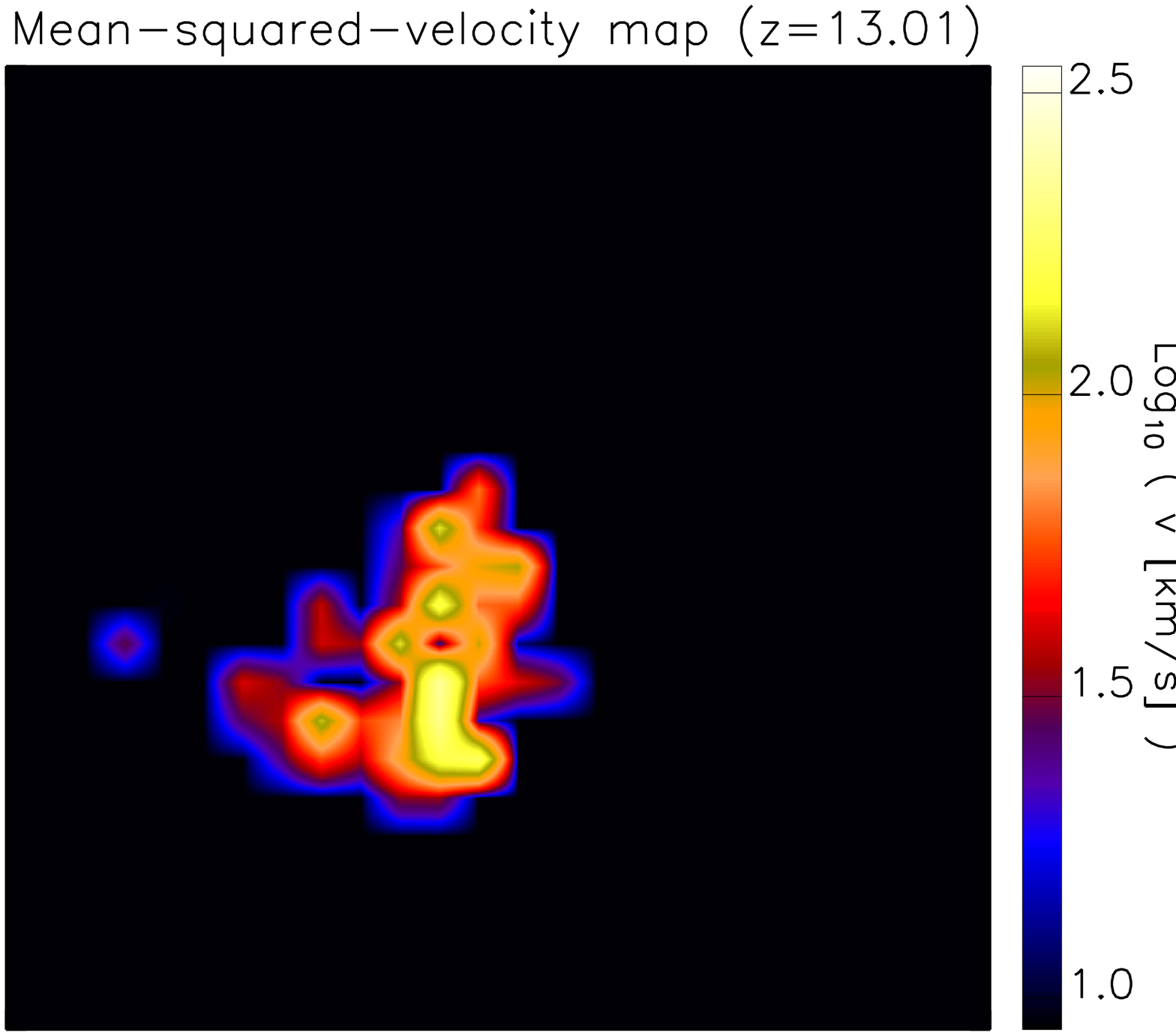}\\
\hspace{1cm}\\
\includegraphics[width=0.29\textwidth]{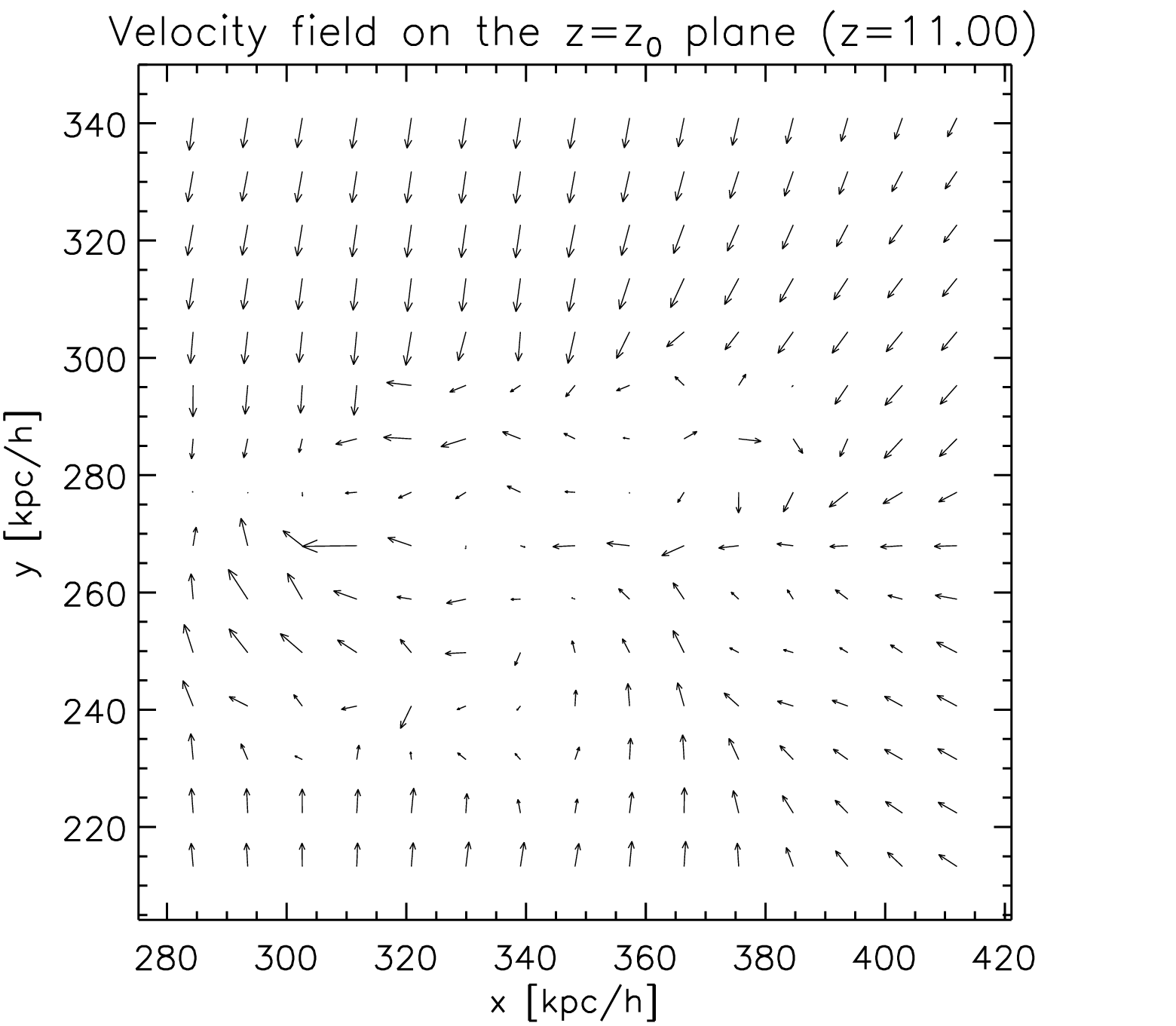}
\includegraphics[width=0.29\textwidth]{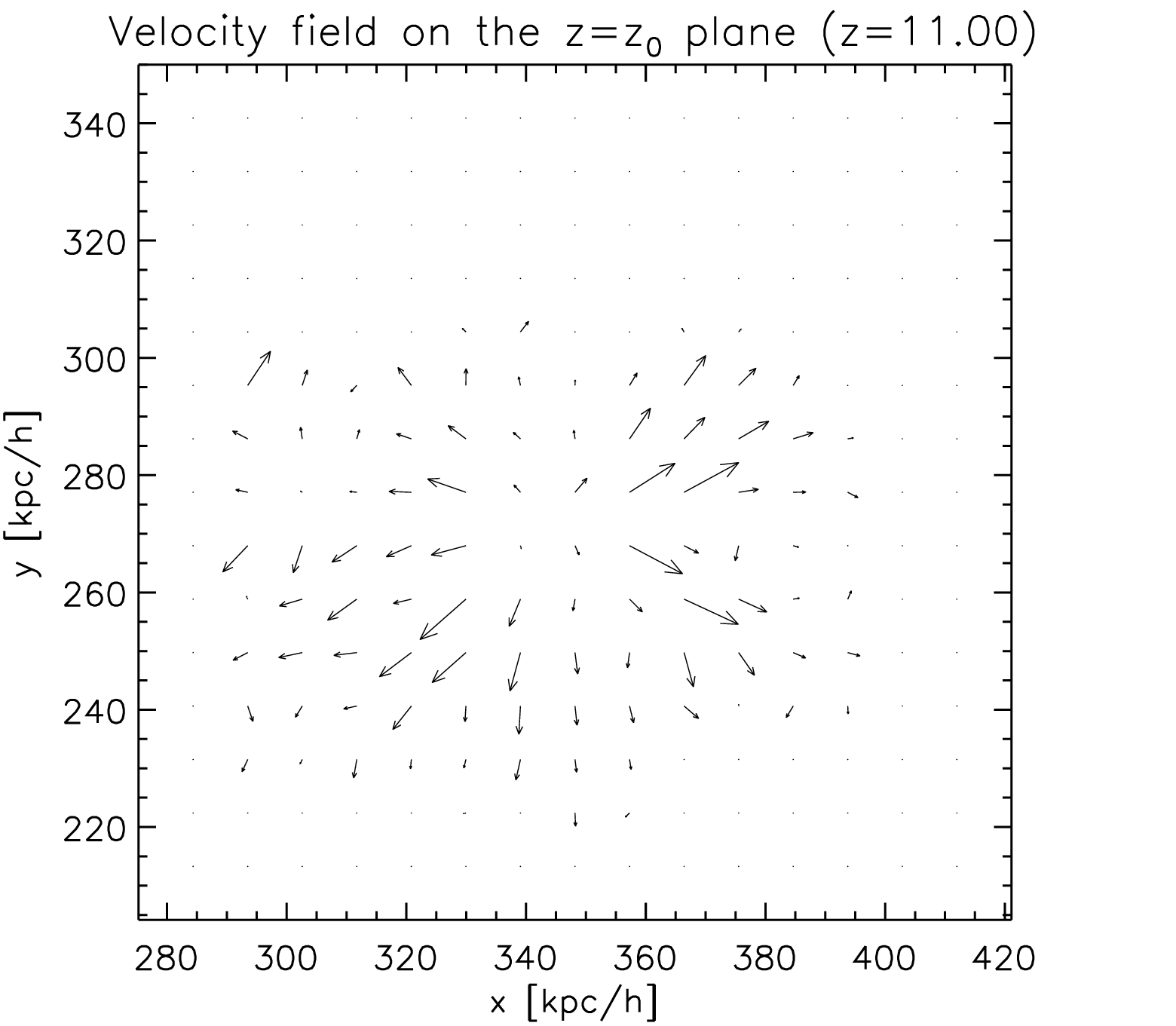}
\includegraphics[width=0.29\textwidth]{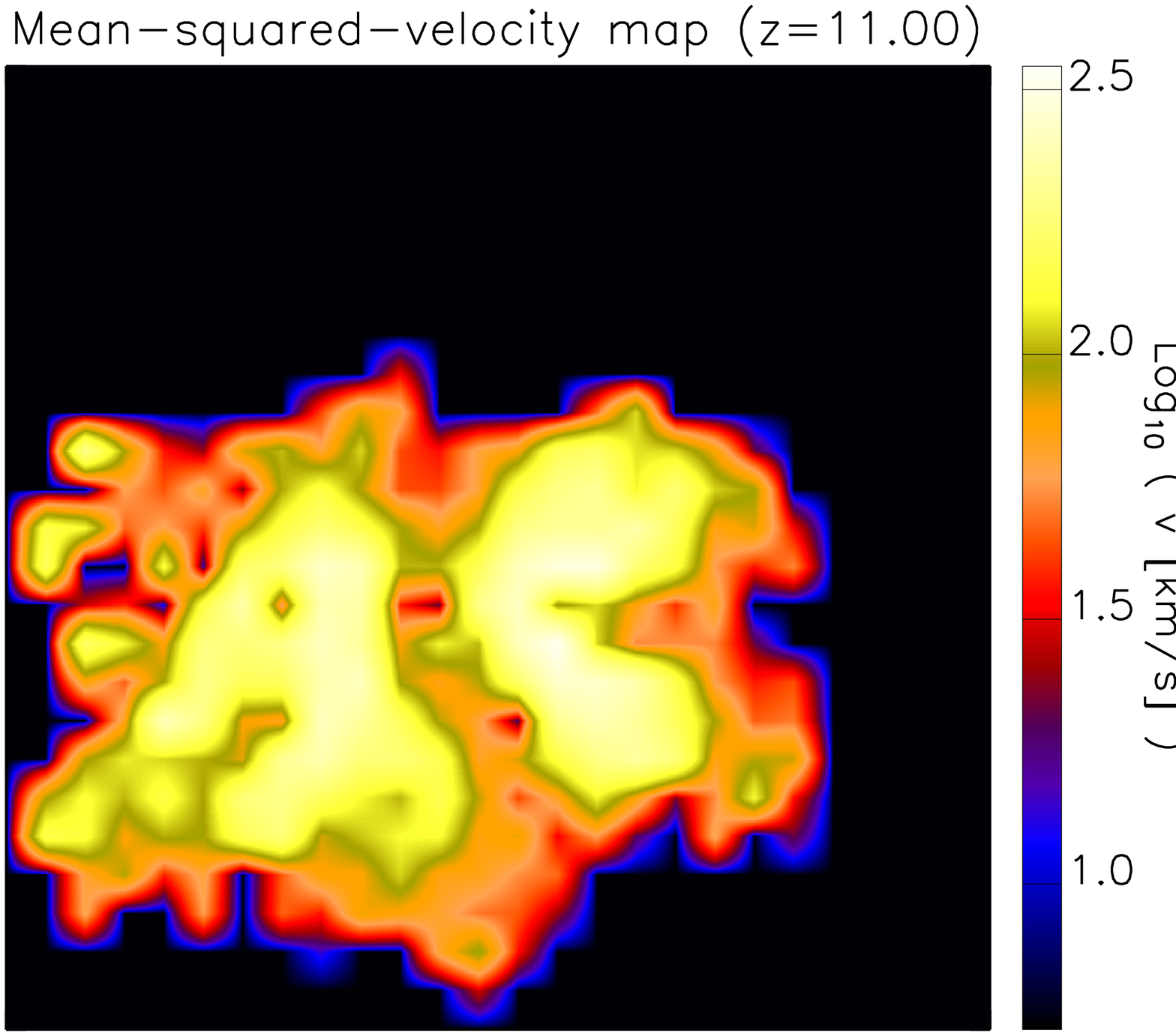}\\
\caption[Velocity fields]{\small
Redshift evolution of the velocity field lines for pristine gas (left column), and metal-rich gas (central column).
Corresponding mean-squared velocity maps for metal-rich gas are shown on the right column.
The selected region is a zoom-in on the plane $z=z_0\simeq 370\,\rm kpc/{\it h}$ (comoving) of the main star forming halo.
}
\label{fig:velocity_field}
\end{figure*}

To get a better understanding of metal pollution at early times, we discuss the dynamical features of the material involved in the spreading process.
As an example, we focus on the evolution of the most massive dark-matter halo (mass between $\sim 10^7-10^8\,\rm M_\odot/{\it h}$ at $z\sim 16-10$) and compute radially averaged quantities.
The comoving virial radius is computed as the distance from the center of mass at which the average enclosed over-density is $\sim 200$, and, at these stages, it varies between $\sim 4-9\,\rm kpc/{\it h}$.
We note that such over-densities include the whole cooling branch and a relevant part of the isothermal gas phase (see e.g. Fig. \ref{fig:PhaseEvolution}).

\subsubsection{Global features}
In Fig. \ref{fig:shells}, we show the mass profiles for metal-enriched and pristine gas of the main halo, at different redshift. The vertical lines refer to the virial radius at each redshift.
Pollution starts at $z\sim 16$, when the early PISN spread the first metals.
From this moment on, star formation contributes more and more  to metal spreading and the ``inside-out'' mode of metal enrichment is visible in the time sequence shown throughout the panels, as more and more surrounding regions are affected by pollution.
The structure of the pristine gas seems strongly influenced by star formation and feedback effects.
In fact, while star formation goes on and the metallicity of the environment increases, the pristine gas is always kept at $\sim 0.2\rm\, kpc$ (comoving): early popIII activity halts it from getting accumulated toward the center.
Only at redshift $z \lesssim 13$, when the residual popIII contribution to the SFR is just $\sim 10^{-3}$, the popII regime dominates, the infall becomes more efficient and this determines the bump at small radii.
The average amount of metal mass in the shells is some orders of magnitude lower than the pristine gas, but locally, it is possible to find highly enriched areas with metallicities up to $Z\sim 10^{-3}$, i.e. $\sim 10^{-1}\,Z_\odot$ (e.g. see at radii of $\sim 0.3\,\rm kpc/{\it h}$, in Fig. \ref{fig:shells}).
Winds are also able to spread around material beyond the virial radius, into nearby empty regions.
\\
This behaviour is confirmed in Fig. \ref{fig:rates}, where we show corresponding mass inflow and outflow rates for metal-rich and pristine gas, in the main halo.
These are computed as
\begin{equation}
\frac{\d M}{\d t} \simeq \frac{\Delta M}{\Delta r} v_{rad},
\end{equation}
with $\Delta M$ mass,
$\Delta r$ the radial bin, and
$v_{rad}$ average radial velocity for the considered shell.
The pristine-gas infall rate is $\sim 10^{-3}-10^{-1}\,\rm M_\odot\, yr^{-1}$.
The metal outflow rates are a few orders of magnitude lower than the infalling rates of pristine gas, but the continuous ejection of metals from PISN/SN pollutes the surroundings up to distances exceeding the virial radius.
This is recovered in the inflow rates of the enriched gas.
While the innermost regions are dominated by chaotic turbulent motions (see following discussion), the outermost enriched mass, situated at distances comparable with the virial radius, is definitely decoupled from the metals in the core and is fastly moving inwards (the radial component of the velocity easily reaches $\sim 50$km/s).
This incoming material has just been pushed away from other sources and undergone wind feedback. It gets into the main halo already at $\sim 16$, and, in the following stages, polluted material systematically replenishes the innermost regions.
Its contribution to the mass profiles becomes comparable or even dominant from $\sim 0.5\,\rm kpc/{\it h}$ (comoving) up to larger radii.
Since wind feedback is the main responsible for metal spreading up to distances larger than a few $\sim\rm kpc$ within $\sim 10^7\,\rm yr$, the presence of metals in these dense environments after very short timescales is a striking sign of pollution from neraby star forming haloes, whose particles are ejected and trapped in the neighbouring collapsed regions.
The whole process, thus, leads to the concept of {\it ``gravitational enrichment''}, i.e. metal enrichment induced by different, external sources, whose polluted material is expelled and re-captured by the gravitational field of closeby haloes, located at some kpc away.
This reminds of genetic enrichment \cite[e.g.][]{Scannapieco_et_al_2003}, but, in this case, there is no obvious parental relationship (child/progenitor haloes) among the involved objects and they can have quite dinstinct evulution history.
In contrast to self-enrichment, that is probably the main way of enriching rare, isolated haloes, up to overcritical $Z$, the gravitational-enrichment mechanism appears to be the dominant one, especially in crowded and actively star forming environments: it can highly boost the metallicities and allow the gas to exceed $Z_{crit}$ in only $\sim 10^7\,\rm yr$ (see further discussion in Sect. \ref{sect:disc}).
\\
In Fig. \ref{fig:velocity_field}, we show velocity maps for enriched and pristine gas, at different redshifts.
The initial collapse of primordial gas appears along the field lines at high $z$ (left column).
This leads to star formation, followed by metal pollution, with consequent spreading of particles, because of wind effects.
Since the typical temperatures of the star forming regions vary between a few hundred Kelvin (for the cold phase) and $\sim 10^6\,\rm K$ (for the gas heated by SN feedback), the local sound speed is $\sim 1-10^2\,\rm km/s$, and, therefore, the star forming environment is characterized by supersonic flows.
Moreover, the interplay between the hot, outflowing, enriched gas (central and right column), and the cold, infalling, pristine gas, leads to the development of a turbulent regime on kpc scales.
To quantify this\footnote{
The effects of numerical viscosity on structure masses and velocy fields are widely discussed in literature. See, e.g., \cite{Dolag_et_al_2005}, \cite{Valdarnini2010arXiv}, or Appendix~A in \cite{Biffi2010arXiv}.
},
we give an order-of-magnitude estimate of the Reynolds number \cite[e.g. see][and references therein]{Elmegreen_Scalo_2004},
\begin{equation}
R \sim ul/\nu \sim {\cal M}\,l\,n\,\sigma,
\end{equation}
with, $u$ velocity, $l$ typical scale of the flux, $\nu$ kinematic viscosity of the gas, $\cal M$ Mach number, $n$ number density, and  $\sigma$ particle cross section.
We have $u\sim 10^2\,\rm km/s$, a fiducial scale -- see Figs. \ref{fig:shells}, and \ref{fig:rates} -- of $l\sim \rm kpc$ (in the simulations the typical dimensions of the star forming regions at the redshifts of interest go up to several comoving kpc), and $\nu\sim c/n\sigma$, where $c$ is the sound speed.
Typical densities of the star forming regions are $n\gtrsim 10^2\,\rm cm^{-3}$, with $\sigma\sim 10^{-15}\,\rm cm^2$ \cite[][]{Elmegreen_Scalo_2004}.
This implies $\nu\sim 10^{18}-10^{20}\,\rm cm^2/s$, and $R\sim 10^8-10^{10}$.
Such high Reynolds numbers arise from the strongly turbulent state of the star forming regions.
These are, of course, order-of-magnitude estimates, but they are solid enough to characterize the nature of such flows.
In addition, the interactions between different gas phases generates strong hydro-instabilities: e.g., Raleigh-Taylor instabilities, originated by the hot, enriched gas pushing against the cold, pristine one in its surroundings (shocks), or Kelvin-Helmholtz instabilities, due to the velocity differences between the outgoing, enriched gas and the infalling, pristine gas.
The interplay among the two phases is very evident in the pristine-gas field lines (left column), compared with the simultaneous metal-gas ones (central column), and with the corresponding velocity maps (right column).
These considerations are expected to hold not only in this particular case, but also for other small primordial haloes, that are highly affected by feedback effects.

\subsubsection{PopIII and  popII}
\begin{figure*}
\centering
\includegraphics[width=0.32\textwidth, height=0.2\textheight]{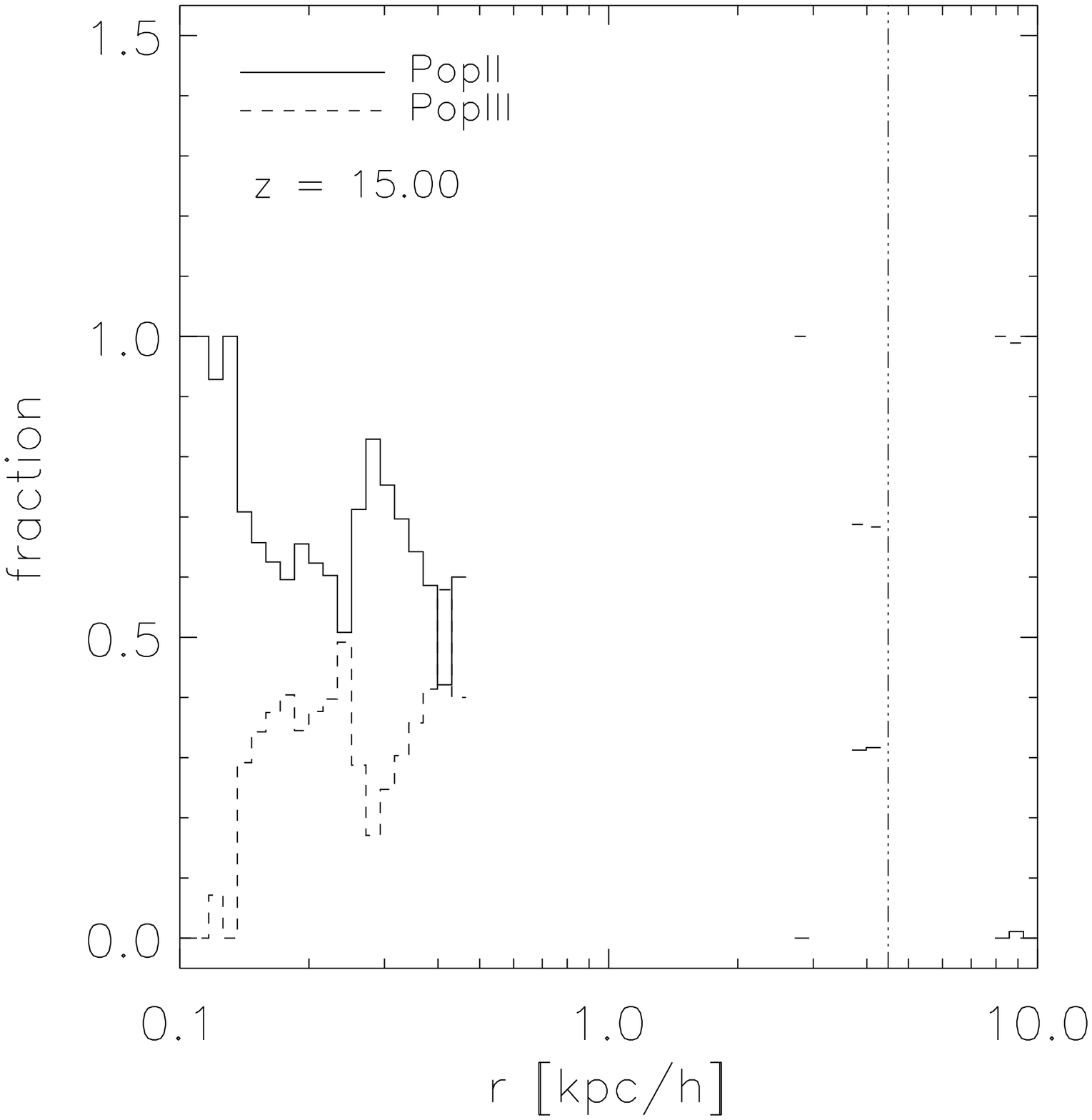}
\includegraphics[width=0.32\textwidth, height=0.2\textheight]{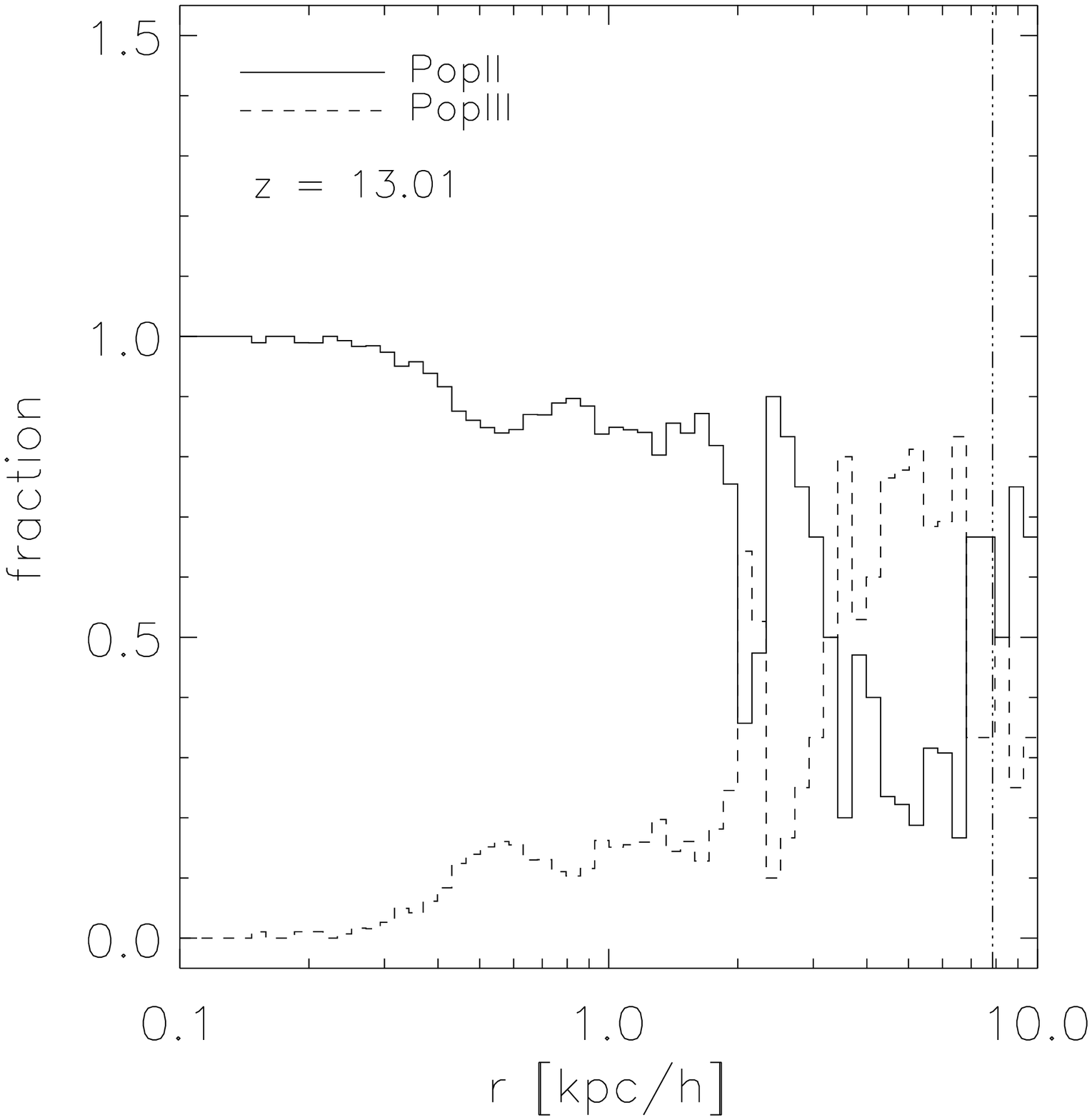}
\includegraphics[width=0.32\textwidth, height=0.2\textheight]{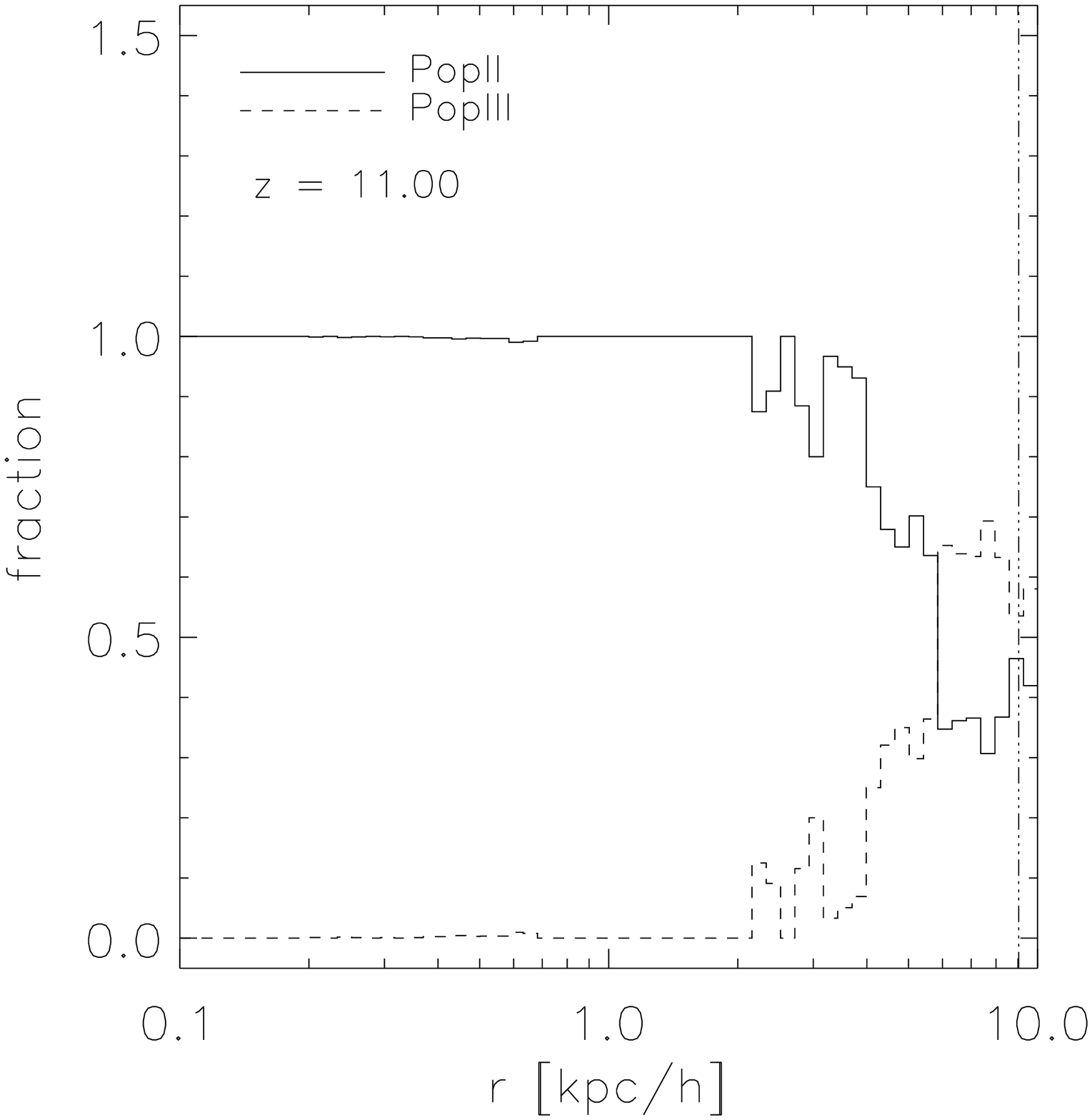}
\caption[Distribution $Z_{crit}$]{\small
Average fractions of popII (solid lines) and popIII (dashed lines) species as a function of comoving radius for the enriched material. The vertical dot-dashed lines indicate the virial radius.
Data refer to  $z=15.00$ (left), $z=13.01$ (center), and $z=11.00$ (right).
}
\label{fig:FlowComposition}
\end{figure*}
In Fig. \ref{fig:FlowComposition}, we summarize our findings about chemical feedback by showing in detail the radial distribution of popIII and popII species, in the innermost regions of the main halo, at redshift $15.00$, $13.01$, and $11.00$.
As time passes, the most striking and evident feature is the gradual disappearance of the popIII generation.
At $z\sim 15$, almost half of the gas is still popIII at comoving distances $\gtrsim 0.2\,\rm kpc/{\it h}$.
Later on, at $z\sim 13$, the situation is similar, but the residual popIII gas is found at larger and larger distances from the center (more than $\sim 2\,\rm kpc/{\it h}$ comoving).
Indeed, new polluted material has been ejected, and, as a consequence, the metallicities of the pre-enriched regions have increased.
So, only the pristine gas far away from star formation sites can maintain a residual metallicity below $Z_{crit}$.
By redshift $\sim 11$, ongoing spreading events will continue to confine the pristine gas to the peripheral regions, several kpc away from the center.
Very similar trends are found when considering separately the infalling portion and the outflowing portion of the gas.


\section{Discussion and conclusions}\label{sect:disc}
We summarize, now, how, throughout this paper, we have addressed the issues presented in the introduction (Sect. \ref{sect:introduction}).
We have studied early structure formation and metal pollution by analyzing numerical N-body/SPH simulations, as described in \cite{Maio2010}.
With respect to other implementations of metal spreading present in literature \cite[e.g.][]{Raiteri1996,Gnedin1998,Mosconi2001,Lia2002a,Lia2002b,KawataGibson2003,Kobayashi2004,RicottiOstriker2004,Cecilia_et_al_2005,TBDM2007,Oppenheimer_et_al_2009, Finlator_et_al_2011},
our implementation additionally follows
proper molecular-chemistry treatment of e$^-$, H, H$^+$, H$^-$, He, He$^+$, He$^{++}$, H$_2$, H$_2^+$, D, D$^+$, HD, and HeH$^+$;
the evolution of individual metal species (C, O, Mg, S, Si, and Fe, since they are the most abundant element produced during stellar evolution);
the transition from the early popIII regime to the subsequent popII regime; and stellar feedback for star forming particles.
\\
Metal enrichment takes place efficiently with run-away molecular cooling leading to the first popIII star formation events, and to the following, rapid transition to popII.
As a consequence of the chemical feedback, higher values of $Z_{crit}$ allow the popIII regime to last a bit longer (at most $<8\times 10^7\,\rm yr$) and to pollute more with respect to the lower-$Z_{crit}$ cases \cite[][]{Maio2010}.
The first sites of metal pollution are clumped and dense star forming regions from which metals are spread away via feedback effects.\\
Because of this process starting in regions with over-densities higher than $\delta\sim 10^4$, the metal filling factor (Sect. \ref{sect:ff}) is extremely small at initial times, and subsequently increases by several orders of magnitude, when metals get into lower-density and larger-volume regions (see Fig. \ref{fig:ff}, left panel).
In roughly $2\times 10^8\,\rm yr$, there is a steep increment of about 8 orders of magnitude, due to the explosions of the early, short-lived, massive stars and the pollution of low-density regions.
Discrimination by $Z_{crit}$ allows us to see that popIII particles dominates the filling factors only at high redshift, and are expected to be found in less dense or more isolated regions at later times.
The precise value of $Z_{crit}$  in the range $\rm [10^{-6},10^{-3}] Z_\odot$ seems to be not crucial for the properties of primordial gas and metal enrichment (Fig. \ref{fig:ff}, right panel), as also argued in \cite{Maio2010}.\\
The phase-space evolution (Fig. \ref{fig:PhaseEvolution}) and the probability distributions (Figs. \ref{fig:GasDistribution}, and \ref{fig:probabilities}) clearly indicate a dominant ``inside-out'' mode of metal pollution, according to which metals are strongly spread from the clumped regions to the low-density regions.
Within a time of $\sim 10^8\,\rm yr$, the newly enriched material accounts for $\sim 90\%$ of the ejected metals (see Sect. \ref{sect:statistics}, where we compute the enrichment rate $\dot\varepsilon_Z\sim 10^{-8}\,\rm yr$).
Moreover, the fraction of enriched material expelled by the winds that falls in the run-away collapsing region and becomes available again for star formation reaches values of $\lesssim 10\%$ (see Sect. \ref{sect:statistics}).
Details about the exact value of $Z_{crit}$ alter only slightly these findings.
In order to see the effects of different popIII IMF's, we have also compared the top-heavy case with the Salpeter case. In the latter, a slightly longer delay of metal enrichment is found, due to the longer stellar lifetimes, but the subsequent pollution history is quite similar \cite[see also][for more details]{Maio2010}.
\\
Our dynamical analysis (e.g.: Figs. \ref{fig:shells}, \ref{fig:rates}, and \ref{fig:velocity_field}) confirms that pristine-gas inflow rates of $\sim 10^{-3}-10^{-1}\,\rm M_\odot\,yr^{-1}$ coexist with metal outflows a few orders of magnitude smaller. Metal-rich winds of hundreds km/s are responsible for pollution of the medium even beyond the virial radii, induce hydro-instabilities, and cause turbulence with Reynolds numbers of $\sim 10^{8}-10^{10}$ (Sect. \ref{sect:dynamics}).
\\
Accumulation of metals in the high-density regions is led by ``gravitational enrichment'' that can boost the metallicity in a few $10^7\,\rm yr$ (see e.g. Figs. \ref{fig:shells} and \ref{fig:rates}), without the need to wait for the growth of the halo to reaccrete them (self-enrichment).
{
We note that alternative scenarios, such as self-enrichment and genetic enrichment, could not explain all the aspects shown by metal pollution at high redshift.
For example, due to the high velocity of the material escaping from the star forming sites, metals can be ejected as far as several kpc away and the halo could effectively re-accrete them only if the growth process were still taking place.
Because this requires at least some $\sim 10^8\,\rm yr$ \cite[e.g.][]{Greif_et_al_2010}, self-enrichment would fail in explaining the high metallicities reached in $\sim 10^7\,\rm yr$, as discussed in Sect.~\ref{sect:dynamics} \cite[see also][]{Maio2010}.
On the other hand, genetic enrichment relies on parental relationships (child/progenitor haloes), and it would not explain metal pollution in haloes which have no direct connections to each other.
\\
More realistically, ``gravitational enrichment'' can account for the pollution of separate haloes which evolve independently.
In fact, when metals are spread beyond the boundaries of a halo (see e.g. previous sections), the gravitational force of a neighboring halo can attract those metals and accrete them onto it.
In this way, metal pollution is enhanced in haloes which have gravitationally trapped metals, originated in completely different environments.
The inside-out mode of metal pollution is at the origin of such dynamical behaviour.
}
\\
Similar results hold for both popII and popIII components (see Fig. \ref{fig:GasDistributionZvsZcrit} and \ref{fig:FlowComposition}).
Due to the previous considerations on metal enrichment, the popIII material is more likely to be found on the borders of primordial haloes or in isolated, low-density environments ($\sim 50\%$ after $\sim 10^8\,\rm yr$), and the popII one in more clustered regions ($\sim 87\%$ after the same time).
A few $10^8\,\rm yr$ after the onset of star formation, the amount of popII material becomes highly dominant over the residual popIII, reaching mass fractions higher than $\sim 99\%$.
\\
One of the most striking conclusions we can draw is related to the rapidity and efficiency by which popII stars are formed after the first bursts of star formation.
Because of the high metal yields of primordial massive PISN/SN and pollution coming from closeby star forming regions, the transition from the popIII to the popII regime is extremely rapid.
Moreover, the mass trends presented in the previous sections (see e.g. Figs. \ref{fig:shells} and \ref{fig:rates}) support this conclusion, as, due to star formation feedback, inflows do not efficiently replenish the star forming regions with newly accreted, pristine gas.
One can estimate to first order the necessary pristine inflow rates, $\dot{M}_{prist}$, needed to dilute metal-rich gas, $M_Z$, to metallicities $Z < Z_{crit}$:
\begin{equation}
\frac{M_Z}{M_Z + M_{prist}} < Z_{crit},
\end{equation}
and
\begin{equation}
M_{prist} > \frac{(1-Z_{crit}) M_Z }{Z_{crit}}.
\end{equation}
That means (for $Z\ll 1$):
\begin{equation}\label{MdotprMdotZ}
\frac{{\dot M_{prist}}}{\dot{M_{Z}}} > \frac{1-Z_{crit}}{ Z_{crit}}\sim \frac{1}{Z_{crit}},
\end{equation}
i.e., for $Z_{crit}\sim 10^{-4}Z_\odot$, one can dilute the ejected metals below $Z_{crit}$ only with $\dot M_{prist}/\dot M_Z\gtrsim 5\times 10^5$.
These huge differences of several orders of magnitude between enriched-gas and pristine-gas rates are usually not reached, as clearly shown in Fig. \ref{fig:rates}.
In general, this is probably what happens in the primordial haloes, which have similar sizes and which experience similar star formation feedback effects.\\
Furthermore, after the onset of star formation, the hot temperatures induced by SN thermal feedback keep dissociating molecules and it becomes difficult for the pristine gas to cool down: in this regime, the metal fine-structure transition cooling plays a dominant role, and that is the reason why one finds enriched cooling gas, ejected by wind feedback from nearby sources, deeper in the potential wells.
Thus, popIII can survive mostly outside clumped regions, in isolated, or in low density environments, where relation (\ref{MdotprMdotZ}) is more easily satisfied, while popII dominates the dense, star forming environments.
\\
We note that the high resolution of our simulations and the detailed chemistry treatment allow us to make firm, consistent, and robust statements on the gas behaviour, from the pristine, molecular stages, to the following star formation phases, and the popIII/popII metal enrichment events.
In particular, the simulations include a complete, self-consistent treatment of primordial gas (dealing with cooling and molecule evolution), which is lacking in other simple approximations adopted in several previous works \cite[e.g.][]{Kobayashi2004,Cecilia_et_al_2005,Tornatore2007,Oppenheimer_et_al_2009,Finlator_et_al_2011}.
Those implementation can be very useful when investigating metal distributions or present-day star formation, but they obviously fail at high redshift, when molecules are the leading coolants, primordial stars form, and the popIII/popII transition takes place.
Consistently with those works, metal enrichment is very patchy and inhomogeneous, with a characteristic ``inside-out'' mode.
This is particularly true for the primordial popIII regime, whose energetics is much more powerful than the popII one.
On the other hand, there are some {\it caveats}: given the small size of the boxes, we are limited in the statistical sampling of rare, large haloes.
In the simulations presented here there is no radiative-transfer treatment from stars, which can affect the very close surroundings of individual massive popIII stars, by heating and ionizing the gas \cite[e.g.][]{ANM1999,Haiman_et_al_2000,Ricotti_et_al_2001,KitayamaYoshida2005,Iliev_et_al_2005,AhnShapiro2007,Wise2008,Greif2009}, but for low-mass SN, this effect would be less severe \cite[e.g.][]{Hasegawa_et_al_2009,Whalen_et_al_2010}.
The inclusion of radiative transfer is relativelly easy, though, only when dealing with individual or a few sources. It becomes very problematic and computationally expensive when studying cosmological evolution with enormous numbers of stars whose radiation has to be followed simultaneously with gas chemistry and hydrodynamics.
However, we do not expect substantial changes on our overall conclusions about the general picture of the popIII/popII transition, or the dynamics of the metals.
Indeed, metals ejected with velocities of $\sim 10^2\,\rm km/s$ would probably be very little affected by radiative pressure \cite[e.g.][]{Johnson_et_al_2010}.
The stellar radiation would probably enhance (see e.g. Fig. \ref{fig:shells}) the evacuation of pristine gas from the individual minihaloes and would contribute to destroy H$_2$ molecules \cite[e.g.][]{OmukaiNishi1999}, leaving, at the same time, the produced metals as only relevant coolants.
This would essentially reproduce the same scenario outlined in Sect. \ref{sect:dynamics}.
Star formation in purely pristine haloes delayed by radiative or thermal feedback would require more mass to be assembled and atomic cooling to become efficient: due to the higher potential wells, metal spreading in such haloes would probably be less efficient.
\\
Metal mixing due to diffusion processes is taken into account by smoothing metallicities on the kernel. This might be a coarse approximation \cite[see also discussion in ][and references therein]{Maio2010}, but realistic, solid and consistent implementations of (large-scale) metal diffusion are currently lacking \cite[approximated attempts are given by, e.g.,][]{Spitzer1962,CowieMcKee1977,Brookshaw1985,Sarazin1988,Monaghan1992,ClearyMonaghan1999,AvillezMacLow2002,KlessenLin2003,Jubelgas_et_al_2004,Monaghan_et_al_2005,Wadsley_et_al_2008,Greif2009,Shen_et_al_2010}.
One should expect that large-scale metal mixing could increase the metal-enriched mass which is spread around or falls back to the high-density environments, but no definitive conclusions can be drawn, at the present stage.
\\
The precise cosmological parameters of the $\Lambda$CDM model adopted could slightly affect the timescale of the picture, and significant changes in the background cosmological frame could have some impact, as structure formation could be significantly shifted in redshift. None the less, the global trends \cite[e.g. see][]{Maio2009} are likely to be recovered.
\\
To conclude, we can state that:
\begin{itemize}
\item
metal pollution is a very efficient process and the critical metallicity is easily reached in less than $\sim 10^8\,\rm yr$;
\item
the population II regime is, in general, the dominant regime of star formation;
\item
the population III regime is expected to survive in isolated environments or in the outskirts of star forming haloes;
\item
metal pollution follows an ``inside-out'' mode and early metals are spread far away from their birth sites, even outside the virial radius;
\item
the inside-out mode leads metal pollution in closeby star forming haloes, inducing ``gravitational enrichment'', while self-enrichment is a negligeble process;
\item
the coexistence of cold, pristine-gas inflows and of hot, enriched-gas outflows determines turbulence with $R\sim 10^8-10^{10}$ and hydro instabilities;
\item
different $Z_{crit}$ or popIII IMF alter only slightly these findings.
\end{itemize}


\section*{acknowledgments}
We warmly acknowledge the referee, K.~Omukai, for his comments on the manuscript.
We also aknowledge useful discussions with R.~Schneider, N.~Yoshida, C.~Dalla~Vecchia, and J.~P.~Paardekooper.
SK acknowledges support from the DFG Cluster of Excellence 'Origins and Structure of the Universe' and from the the Royal Society Joint Projects Grant JP0869822.


%
%


\bibliographystyle{mn2e}
\bibliography{bibl}

\label{lastpage}
\end{document}